\documentclass{article}
\usepackage{graphicx} 
\usepackage{amsmath}
\usepackage{amssymb}
\usepackage{amsthm}
\usepackage{thmtools}
\usepackage{thm-restate}
\usepackage{bm}
\usepackage{graphicx}
\usepackage{caption}
\usepackage[margin = 1in]{geometry}
\usepackage{natbib}
\usepackage[normalem]{ulem}
\setcitestyle{numbers,square}
\usepackage{color}
\usepackage[dvipsnames]{xcolor}
\usepackage{media9}
\usepackage{subcaption}
\usepackage{booktabs}
\usepackage{comment}
\usepackage{enumitem}
\usepackage[colorlinks,urlcolor=blue,citecolor=ForestGreen,linkcolor=ForestGreen]{hyperref}
\usepackage[nameinlink,capitalize,nosort]{cleveref}
\crefname{claim}{Claim}{Claims}
\usepackage{marginnote}
\usepackage{tikz}
\usetikzlibrary{decorations.pathreplacing}

\usepackage{algorithm}
\usepackage[noend]{algpseudocode}

\usepackage[colorinlistoftodos]{todonotes}
\usepackage[most]{tcolorbox}

\usepackage[utf8]{inputenc}

\definecolor{mygreen}{rgb}{0,0.65,0.66}

\newcommand{\poly}{\text{poly}}
\newcommand{\polylog}{\text{polylog}}

\graphicspath{{imgs/}}

\makeatletter
\def\mathcolor#1#{\@mathcolor{#1}}
\def\@mathcolor#1#2#3{%
  \protect\leavevmode
  \begingroup
    \color#1{#2}#3%
  \endgroup
}
\makeatother

\newcommand*{\vsepfbox}[1]{%
  \begingroup
    \sbox0{\fbox{#1}}%
    \setlength{\fboxrule}{0pt}%
    \mbox{\kern-\fboxsep\fbox{\unhbox0}\kern-\fboxsep}%
  \endgroup
}

\theoremstyle{plain} \numberwithin{equation}{section}
\newtheorem{theorem}{Theorem}[section]
\numberwithin{theorem}{section}

\newtheorem{lemma}[theorem]{Lemma}

\newtheorem{definition}[theorem]{Definition}

\theoremstyle{definition}

\usepackage{thmtools} 
\usepackage{thm-restate}

\DeclareMathOperator*{\argmin}{argmin}

\makeatletter
\newcommand{\subalign}[1]{%
  \vcenter{%
    \Let@ \restore@math@cr \default@tag
    \baselineskip\fontdimen10 \scriptfont\tw@
    \advance\baselineskip\fontdimen12 \scriptfont\tw@
    \lineskip\thr@@\fontdimen8 \scriptfont\thr@@
    \lineskiplimit\lineskip
    \ialign{\hfil$\m@th\scriptstyle##$&$\m@th\scriptstyle{}##$\hfil\crcr
      #1\crcr
    }%
  }%
}
\makeatother

\newcounter{note}[section]

\newcommand{\lpvalue}{X_H}

\newcommand{\dalg}{d_{T}}
\newcommand{\numlevels}{\log{k}}

\title{A Little Length Slack Goes a Long Way in\\Planar Length-Constrained Network Design}

\author{D Ellis Hershkowitz\thanks{Brown University. Supported in part by NSF grant CCF-2403236 and NSF CAREER Award CCF-2540331.} \and Richard Z Huang\footnotemark[1] \and Rhea Jain\thanks{Carnegie Mellon University. Supported by postdoctoral fellowship made possible by the Vasantrao Shrinivas Dempo Reflective Chair awarded to Prof. R Ravi.} \and Manas Korimilli\footnotemark[1]}

\date{}
\begin{document}

\maketitle
\begin{abstract}
Problems in length-constrained network design ask for min-cost subgraphs connecting $k$ input vertex pairs by paths of length at most $h$. Algorithms that instead connect these pairs by paths of length at most $\beta h$ are said to have length slack $\beta$. Recent work has shown that poly-time polylogarithmic approximations with polylogarithmic length slack are possible for a broad class of both directed and multicommodity length-constrained network design problems on planar graphs.

We give the first poly-time polylogarithmic approximations with $O(1)$ length slack for these problems on planar graphs. In particular, for the length-constrained setting, we give LP-competitive algorithms for planar graphs with length slack $1+\epsilon$ and approximation ratios of $O(\log^2 k/\epsilon)$ for directed Steiner tree and $O(\log^5 k/\epsilon^2)$ for (undirected) Steiner forest. Similarly, for length-constrained directed Steiner forest, we give an $O(\log^6 k/\epsilon)$ approximation on planar graphs with length slack $6+\epsilon$. Our approach is based on a novel charging argument that charges the cost of our solution to the length of the LP and the length of our solution to the cost of the LP. We complement these algorithms with hardness results that separate planar and general graphs for both forest problems: under standard complexity assumptions, we rule out polylogarithmic approximations on general graphs with length slack below $3$.
\end{abstract}
\thispagestyle{empty}
\newpage
\pagenumbering{gobble}
\newpage

\pagenumbering{arabic}

\setcounter{page}{1}

\section{Introduction}
Network design is an area of combinatorial optimization concerned with efficiently constructing networks that connect specified nodes at low cost. The subarea of \emph{length-constrained network design} additionally requires that these nodes be connected by \emph{short} paths \cite{marathe1998bicriteria,babay2018lcsf,hershkowitz2025simple,chekuri2026lcdsf}. Such problems arise in numerous practical settings, including transportation networks, where depots must be connected by short routes, and communication networks, where messages must be transmitted with low latency \cite{akgun2011new}. 

The general setup for length-constrained problems is as follows: given a graph $G=(V,E)$ with nonnegative edge cost and length functions $c,l:E\to\mathbb{Z}_{\geq0}$, a connectivity requirement, and an integer length constraint $h>0$, output a minimum cost subgraph satisfying the connectivity requirement over paths of length at most $h$. Different connectivity requirements give different problems. In \textbf{length-constrained Steiner tree (LC-ST)}, we are given the above input along with a root $r \in V$, a set of $k$ terminals $S\subseteq V$, and must output a minimum cost subgraph containing an $r$-$v$ path of length at most $h$ for every $v\in S$. In \textbf{length-constrained Steiner forest (LC-SF)}, we are given the above input along with a set of $k$ terminal pairs $D=\left\{\left(s_i,t_i\right)\right\}_{i \in [k]}$, and must output a minimum cost subgraph containing an $s_i$-$t_i$ path of length at most $h$ for every $i \in [k]$. In other words, LC-ST is a \textit{single-source} problem while LC-SF is its \textit{multicommodity} generalization.\footnote{LC-ST is the special case of LC-SF where $s_i=r$ for every $i\in [k]$.} The directed versions of these problems---\textbf{length-constrained directed Steiner tree (LC-DST)} and \textbf{length-constrained directed Steiner forest (LC-DSF)}---are defined analogously but on directed graphs, and require a directed $r$-$v$ path for every terminal $v$ (LC-DST) and a directed $s_i$-$t_i$ path for every terminal pair (LC-DSF) respectively.

Length-constrained problems are much harder than their non-length-constrained counterparts. For example, (non-length-constrained) Steiner tree and forest both admit poly-time $O(1)$-approximation algorithms and have $O(1)$ integrality gaps for their natural linear programming (LP) relaxations \cite{agrawal1991treesf,byrka2013steinertree,ahmadi2025breakingsf,gupta2026steinerforest}. Meanwhile, LC-ST and LC-SF are equivalent to directed Steiner tree  and directed Steiner forest and so inherit a polylogarithmic and $2^{\log^{1-\epsilon} n}$ hardness of approximation, respectively \cite{halperin2003polylogarithmic,dinitz2015label}. Length constraints also make natural metric \cite{kou1981steinertree,gupta2015glutton} and LP approaches \cite{byrka2013steinertree,agrawal1991treesf} difficult to use.
In particular, the shortest path distances for non-length-constrained problems form a metric, but the natural notion of ``length-constrained'' distances\footnote{The $h$-length-constrained distance between vertices $u,v$ is the cost of the minimum-cost $u,v$ path of length at most $h$.} is inapproximable by any metric \cite{haeupler2021treeemb}. Furthermore, the natural linear program for LC-ST (and hence LC-SF) has an integrality gap of $\Omega(n^\epsilon)$ \cite{li2022flowdst,ZosinK02}. 

However, if we settle for \textit{bicriteria} approximation algorithms, then better poly-time approximations are possible. An \textbf{$\alpha$ approximation algorithm with $\beta$ length slack} is one that outputs a subgraph whose cost is an $\alpha$ approximation of the optimal solution and whose length violates the length constraint by at most a multiplicative $\beta$.\footnote{Importantly, the cost is approximate with respect to the optimal solution with $h$ as the length constraint, not $\beta h$.} Polylogarithmic approximations are known for LC-ST and LC-SF but only at the cost of polylogarithmic length slack \cite{marathe1998bicriteria,chekuri2010approximation,haeupler2021treeemb,filtser2022hop}.
In fact, on general graphs, no polylogarithmic approximations with $O(1)$ length slack are known for any of these problems.  Indeed, known approaches seem to necessarily require polylogarithmic length slack. In particular, all known approaches either build connections over logarithmically many rounds, accumulating $O(h)$ length per round \cite{marathe1998bicriteria}, or inherit the provably necessary polylogarithmic length slack of length-constrained tree embeddings \cite{haeupler2021treeemb,filtser2022hop,fletcher2026spanning}. 



A recent line of work goes beyond polylogarithmic length slack for these problems by studying \textit{planar graphs}. \citet{hershkowitz2026planar} gave a $O(\log^{1+\epsilon}k)$ approximation with $1+\epsilon$ length slack for LC-ST on planar graphs by using a different approach than the above two. However, even on planar graphs, directed and multicommodity problems have proven more resistant to sub-polylogarithmic length-slack. For example, \citet{chekuri2026lcdsf} later gave polylogarithmic approximations for planar LC-DST and LC-DSF, but all of these algorithms had $O(\log k)$ length slack. This slack arises from the aforementioned accumulation of $O(h)$ length per round over $O(\log k)$ recursive rounds. Achieving constant length slack for directed and multicommodity problems therefore seems to require new tools that avoid this accumulation. Given that no polylogarithmic approximation with $O(1)$ length slack is known for the directed or multicommodity versions of any of these problems, we ask:
\begin{quote}\centering
    \textit{Do any directed or multicommodity length-constrained problems admit\\ a polylogarithmic approximation with $O(1)$ length slack on planar graphs?}
\end{quote}
\noindent
At the same time, while some progress has been made on $O(1)$ length slack, we still have a poor understanding of the integrality gaps of natural LP relaxations for planar length-constrained problems. Specifically, in addition to the above result,
\citet{hershkowitz2026planar} also gave an $O(\log^3k/\epsilon)$ LP-competitive approximation with $1+\epsilon$ length slack for LC-ST, and \citet{chekuri2026lcdsf} gave an $O(\log^2 k)$ LP-competitive approximation with length slack $O(\log k)$ for LC-DST where, again, the $O(\log k)$ length slack is lost to $O(h)$ accumulation in each of $O(\log k)$ rounds. Thus, no polylogarithmic LP-competitive approximations are known for LC-DST or LC-SF with $O(1)$ length slack, leading us to ask:
\begin{quote}\centering
    \textit{Do any directed or multicommodity length-constrained problems admit\\ an \textbf{LP-competitive} polylogarithmic approximation with $O(1)$ length slack on planar graphs?}
\end{quote}

\begin{table}[t]
    \renewcommand{\arraystretch}{1.25}
    \begin{tabular*}{\textwidth}{@{\extracolsep{\fill}}lcccc@{}}
        \toprule
        \textbf{Problem} & \textbf{Reference} & \textbf{Cost Approximation} & \textbf{Length Slack} & \textbf{LP Competitive?} \\
        \midrule
        \textbf{LC-DST}
            & \cite{chekuri2026lcdsf} & $O(\log k)$ & $O(\log k)$ & No \\
            & \cite{chekuri2026lcdsf} & $O(\log^2 k)$ & $O(\log k)$ & Yes \\
            & \textbf{This work} (\ref{thm:LCDST}) & $\bm{O(\log^2 k/\epsilon)}$ & $\bm{1+\epsilon}$ & \textbf{Yes} \\
        \midrule
        \textbf{LC-SF}
            & \cite{chekuri2026lcdsf} (directed result) & $O(\log^6 k)$ & $O(\log k)$ & No \\
            & \textbf{This work} (\ref{thm:LCSF}) & $\bm{O(\log^5 k/\epsilon^2)}$ & $\bm{1+\epsilon}$ & \textbf{Yes} \\
        \midrule
        \textbf{LC-DSF}
            & \cite{chekuri2026lcdsf} & $O(\log^6 k)$ & $O(\log k)$ & No \\
            & \textbf{This work} (\ref{thm:LCDSF}) & $\bm{O(\log^6 k/\epsilon)}$ & $\bm{6+\epsilon}$ & \textbf{No} \\
        \bottomrule
    \end{tabular*}
    \caption{Summary of poly-time approximations for length-constrained network design in planar graphs.}\label{tab:planar-results}
\end{table}


\subsection{Our Contributions}

We settle the above questions. First, we give polylogarithmic approximations for LC-DST, LC-SF and LC-DSF, all with $O(1)$ length slack---in the case of LC-DST and LC-SF we even achieve $1+\epsilon$ length slack. Second, our algorithms for LC-DST and LC-SF are LP-competitive, and thus give an upper bound on the corresponding integrality gaps. We summarize our results in \Cref{tab:planar-results}.

Our algorithm for planar LC-DST is an $O\left(\frac{\log^2k}{\epsilon}\right)$ LP-competitive approximation with $1+\epsilon$ length slack.
    \begin{restatable}[LC-DST]{theorem}{thmLCDST}\label{thm:LCDST}
        For any $\epsilon\in(0,1)$, there is an LP-competitive $O\left(\frac{\log^2k}{\epsilon}\right)$ approximation with $1+\epsilon$ length slack for LC-DST in planar graphs that runs in time $\poly(n,1/\epsilon)$.
    \end{restatable}
\noindent
This constitutes a considerable improvement over two of the aforementioned results. First, it improves the cost approximation of the LP-competitive planar LC-ST algorithm of \citet{hershkowitz2026planar}  from $O(\log^3 k/\epsilon)$ to $O(\log^2 k/\epsilon)$---notably, this previous cost bound was only for undirected graphs whereas our algorithm is for directed graphs. Second, it improves the length slack of the planar LC-DST algorithm of \citet{chekuri2026lcdsf} from $O(\log k)$ to $(1+\epsilon)$ with the same LP-competitive cost.

We then use our LC-DST algorithm as a starting point towards building our LC-SF algorithm. In particular, we give an LP-competitive $O\left(\frac{\log^5k}{\epsilon^2}\right)$ approximation for LC-SF in planar graphs with length slack $1+\epsilon$.
    \begin{restatable}[LC-SF]{theorem}{thmLCSF}\label{thm:LCSF}
        For any $\epsilon\in(0,1)$, there is an LP-competitive $O\left(\frac{\log^5k}{\epsilon^2}\right)$  approximation with $1+\epsilon$ length slack for LC-SF in planar graphs that runs in time $\poly(n,1/\epsilon)$. 
    \end{restatable}
\noindent LC-SF was previously not studied specifically in planar graphs. Therefore, the previous best approximations for planar LC-SF were either via (1) LC-SF in general graphs, or (2) LC-DSF in planar graphs. The best known approximations for LC-SF in general graphs lose polylogarithmic factors in both approximation ratio and length slack \cite{haeupler2021treeemb,filtser2022hop}, while the best approximation for planar LC-DSF is a (non-LP-competitive) $O(\log^6k)$ approximation with $O(\log k)$ length slack of \citet{chekuri2026lcdsf}. Our result improves on both. Compared to the planar LC-DSF result, for example, our algorithm is LP-competitive, reduces the cost from $O(\log^6k)$ to $O(\log ^ 5 k)$ and reduces the length slack from $O(\log k)$ to $1+\epsilon$.
    
Lastly, plugging \Cref{thm:LCDST} into an algorithm of \citet{chekuri2026lcdsf} gives our $O\left(\frac{\log^6k}{\epsilon}\right)$ approximation for LC-DSF with length slack $O(1)$.
    \begin{restatable}[LC-DSF]{theorem}{thmLCDSF}\label{thm:LCDSF}
         For any $\epsilon\in(0,1)$, there is an $O\left(\frac{\log^6k}{\epsilon}\right)$ approximation with $6+\epsilon$ length slack for LC-DSF in planar graphs that runs in time $\poly(n,1/\epsilon)$. 
    \end{restatable}
\noindent
Notably, the above improves the length slack of \citet{chekuri2026lcdsf} from $O(\log k)$ to $6+\epsilon$ with no loss in the cost. 

We complement our planar graph algorithms with hardness of approximations for LC-SF and LC-DSF in general graphs. In particular, we show that polylogarithmic LC-SF is impossible with length slack less than $3$ and polylogarithmic LC-DSF is impossible with any amount of length slack.
    \begin{restatable}[Hardness]{theorem}{thmHard}\label{thm:hard}
        For every fixed $\epsilon>0$, there is no polynomial time $2^{\log^{1 - \epsilon}{k}}$ approximation with length slack $\beta$ for $\beta<3$ for LC-SF assuming $\textsf{NP}\not\subseteq \textsf{BPTIME} \left(2^{\polylog(n)}\right)$. For LC-DSF the same statement holds but for any positive length slack $\beta$.
    \end{restatable}
\noindent Taken together with our polylogarithmic approximations for LC-SF with $1+\epsilon$ length slack and LC-DSF with $O(1)$ length slack, the above hardness shows that LC-SF and LC-DSF are provably easier in planar graphs than in general graphs.

\subsection{Overview and Intuition} 
\label{sec:tech_overview}
Our approach is based on several new tools for planar length-constrained network design with $1+\epsilon$ length slack that relies on ``shortcutting'' planar separators.

\begin{figure}[t]
    \centering
\begingroup%
\pgfmathsetlengthmacro{\separatorxunit}{(\linewidth-12pt)/16}%
\definecolor{separatorred}{RGB}{205,55,45}%
\begin{tikzpicture}[
  x=\separatorxunit,y=1cm,
  line cap=round,line join=round,
  graph edge/.style={draw=black,line width=1.2pt},
  separator edge/.style={draw=separatorred,line width=1.7pt},
  attachment/.style={draw=black!65,line width=.85pt,dash pattern=on 3pt off 2pt},
  boundary/.style={draw=black!55,line width=.75pt},
  vertex/.style={circle,inner sep=0pt,outer sep=0pt,minimum size=5pt},
  every node/.style={font=\normalfont\normalsize,inner sep=1pt}
]
  \newcommand{\separatorcomponent}[9]{%
    \begin{scope}[shift={(#3,#4)},xscale=#5,yscale=#6]
      \draw[boundary]
        (-.98,-.10)
        .. controls (-1.02,.50) and (-.70,1.05) .. (-.10,1.02)
        .. controls (.45,1.12) and (1.03,.62) .. (.98,-.03)
        .. controls (1.03,-.50) and (.48,-.81) .. (-.06,-.73)
        .. controls (-.65,-.85) and (-.95,-.55) .. cycle;

      \ifcase#2\relax
      \or
        \coordinate (#1-r) at (-.05,.42);
        \coordinate (#1-a) at (-.40,.02);
        \coordinate (#1-b) at (-.64,-.45);
        \coordinate (#1-c) at (.27,-.02);
        \coordinate (#1-d) at (.62,-.48);
        \coordinate (#1-e) at (.02,-.52);
        \coordinate (#1-f) at (.72,.25);
        \def\componenttreeedges{r/a,r/c}
        \def\componentotheredges{a/b,c/d,c/e,r/f,f/d}
        \def\componenttreevertices{r,a,c}
        \def\componentothervertices{b,d,e,f}
      \or
        \coordinate (#1-r) at (0,.40);
        \coordinate (#1-a) at (-.52,-.08);
        \coordinate (#1-b) at (.03,-.18);
        \coordinate (#1-c) at (.58,.02);
        \coordinate (#1-d) at (.02,-.53);
        \coordinate (#1-e) at (-.57,-.48);
        \coordinate (#1-f) at (.64,-.45);
        \def\componenttreeedges{r/a,r/b}
        \def\componentotheredges{b/d,r/c,a/e,e/d,c/f,f/d}
        \def\componenttreevertices{r,a,b}
        \def\componentothervertices{c,d,e,f}
      \or
        \coordinate (#1-r) at (.05,.44);
        \coordinate (#1-a) at (-.16,.10);
        \coordinate (#1-b) at (-.57,-.24);
        \coordinate (#1-c) at (.20,-.08);
        \coordinate (#1-d) at (.52,-.43);
        \coordinate (#1-e) at (-.13,-.46);
        \coordinate (#1-f) at (.54,.24);
        \coordinate (#1-g) at (.64,-.12);
        \def\componenttreeedges{r/a,r/f}
        \def\componentotheredges{a/b,a/c,c/d,a/e,f/g,g/d,c/g}
        \def\componenttreevertices{r,a,f}
        \def\componentothervertices{b,c,d,e,g}
      \or
        \coordinate (#1-r) at (.23,.44);
        \coordinate (#1-a) at (-.05,.06);
        \coordinate (#1-b) at (-.58,-.20);
        \coordinate (#1-c) at (.05,-.48);
        \coordinate (#1-d) at (.65,-.15);
        \coordinate (#1-e) at (-.57,-.55);
        \def\componenttreeedges{r/a,r/d}
        \def\componentotheredges{a/b,a/c,b/e,e/c}
        \def\componenttreevertices{r,a,d}
        \def\componentothervertices{b,c,e}
      \fi
      \foreach \p/\q in \componentotheredges
        \draw[graph edge] (#1-\p) -- (#1-\q);
      \foreach \p/\q in \componenttreeedges
        \draw[draw=#7,line width=#8] (#1-\p) -- (#1-\q);
      \foreach \p in \componenttreevertices
        \node[vertex,fill=#7] at (#1-\p) {};
      \foreach \p in \componentothervertices
        \node[vertex,fill=black] at (#1-\p) {};
      \node[anchor=base] at (-.50,.40) {$G_{#2}$};
    \end{scope}%
    \ifnum#9=1
      \node[anchor=base] at ([yshift=7pt]#1-r) {$r_{#2}$};
    \fi
  }

  \separatorcomponent{leftone}{1}{.60}{.10}{1.10}{.72}{black}{1.2pt}{0}
  \separatorcomponent{lefttwo}{2}{2.30}{-1.45}{.85}{.80}{black}{1.2pt}{0}
  \separatorcomponent{leftthree}{3}{4.65}{-1.50}{1.05}{.72}{black}{1.2pt}{0}
  \separatorcomponent{leftfour}{4}{6.30}{0}{.90}{.84}{black}{1.2pt}{0}

  \coordinate (r) at (3.10,2);
  \coordinate (qonea) at (2.25,1.10);
  \coordinate (qoneb) at (2.05,.30);
  \coordinate (qonec) at (1.90,-.45);
  \coordinate (shared) at (3.80,1.20);
  \coordinate (qtwoa) at (3.70,.45);
  \coordinate (qtwob) at (3.85,-.45);
  \coordinate (qthreea) at (5.05,.80);
  \coordinate (qthreeb) at (5.10,-.20);

  \foreach \p/\q in {qonea/leftone-f,qoneb/leftone-d,qoneb/lefttwo-r,
                      qtwob/leftthree-r,qthreeb/leftthree-f,
                      qthreeb/leftfour-b}
    \draw[attachment] (\p) -- (\q);
  \foreach \p in {leftone-f,leftone-d,lefttwo-r,
                   leftthree-r,leftthree-f,leftfour-b}
    \node[vertex,fill=black] at (\p) {};

  \draw[separator edge] (r) -- (qonea) -- (qoneb) -- (qonec);
  \draw[separator edge] (r) -- (shared) -- (qtwoa) -- (qtwob);
  \draw[separator edge] (shared) -- (qthreea) -- (qthreeb);
  \foreach \p in {r,qonea,qoneb,qonec,shared,qtwoa,qtwob,qthreea,qthreeb}
    \node[vertex,fill=separatorred] at (\p) {};
  \node[anchor=base] at ([yshift=7pt]r) {$r$};
  \node[text=separatorred,anchor=base east] at (1.95,1.25) {$Q_1$};
  \node[text=separatorred,anchor=base east] at (3.40,.45) {$Q_2$};
  \node[text=separatorred,anchor=base] at (4.70,1.08) {$Q_3$};

  \draw[->,>=stealth,line width=1.1pt] (7.55,.05) -- (8.95,.05);

  \separatorcomponent{rightone}{1}{10.51}{1.04}{1.30}{1.05}{separatorred}{1.7pt}{1}
  \separatorcomponent{righttwo}{2}{13.64}{.99}{1.25}{1.05}{separatorred}{1.7pt}{1}
  \separatorcomponent{rightthree}{3}{10.81}{-1.14}{1.30}{1.05}{separatorred}{1.7pt}{1}
  \separatorcomponent{rightfour}{4}{13.89}{-1.19}{1.22}{1.05}{separatorred}{1.7pt}{1}

  \node[anchor=base] at (3.40,-2.70) {1. Find a separator};
  \node[anchor=base] at (12.30,-2.70) {2. Delete and recurse};
\end{tikzpicture}%
\endgroup%
    \caption{Shortest-path separators (red) in initial graph and in resulting components $G_i$ with new roots $r_i$.}
    \label{fig:separator}
\end{figure}
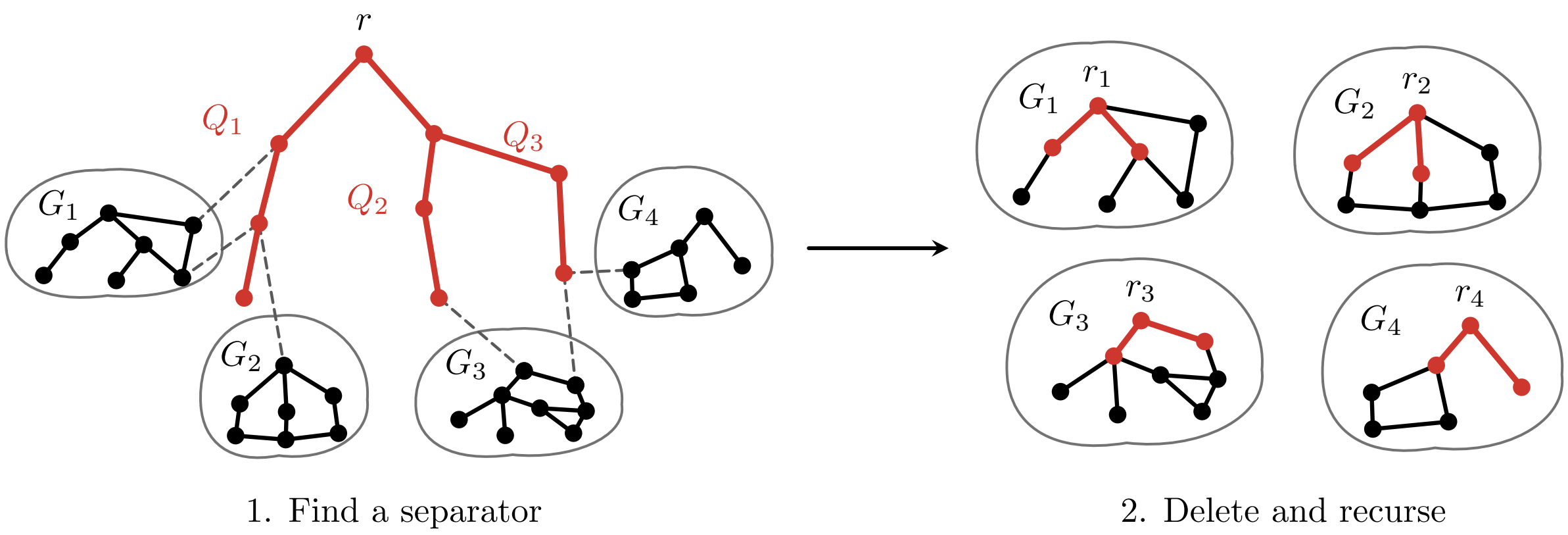

Specifically, much of the recent progress on network design problems in planar graphs \cite{kawarabayashi2022embeddings,friggstad2023dst,chekuri2024dpst,chekuri2025dsf,hershkowitz2026planar,chekuri2026lcdsf} is based on the existence of \emph{shortest path separators} \cite{thorup2004compact,lipton1979separator}. Informally, every planar graph $G = (V, E)$ has three shortest paths from a root $r \in V$ whose removal results in connected components each containing at most a constant fraction of the vertices. These separators allow for divide-and-conquer algorithms, in which one computes and buys a separator, and then recursively solves the problem on each connected component; see \Cref{fig:separator}. In length-constrained settings, one can compute shortest path separators under a metric (known as the \emph{mixture metric}) that averages cost and length functions, yielding separators that have bounded cost and length. These costs and lengths accumulate over the recursive levels, naturally resulting in $O(\log k)$ loss in both approximation ratio and length slack. 

\citet{hershkowitz2026planar} provide an initial framework for improving length slack for planar network design problems by obtaining a polylogarithmic approximation with $1+\epsilon$ length slack for undirected LC-ST. The idea is to partition separators into smaller pieces, and buy ``shortcut'' paths to pieces in order to avoid traversing the full separator. As we discuss below, there are several challenges with directly extending these ideas to directed and multicommodity settings. We develop new tools for planar length-constrained network design with $1+\epsilon$ slack in these settings, which we believe may be of independent interest for other related problems. Our techniques are LP-based, and thus also yield upper bounds on the integrality gaps of natural LP relaxations. 


\subsubsection{Length-Constrained Directed Steiner Tree}
Our LC-DST algorithm is based on a novel analysis which charges the cost of our solution to the length of the LP and the ($(1+\epsilon)h$) length of our solution to the cost of the LP. 

Like in earlier work, we start by computing a \textit{separator hierarchy}. The separator hierarchy is essentially a rooted tree of height $\log k$ where each node corresponds to a region of $G$, its children in the tree correspond to the regions obtained by recursively computing a separator on it, and the regions at each level of the tree form a disjoint union. Hence the root is $G$ itself, and any path from the root in the hierarchy tree is a chain of regions. Thus, every node on the hierarchy has its own separator and a corresponding \textit{parent separator}.

\begin{figure}[h]
    \centering
    \begin{subfigure}[t]{0.5\textwidth}
        \centering
        \includegraphics[width=.5\linewidth]{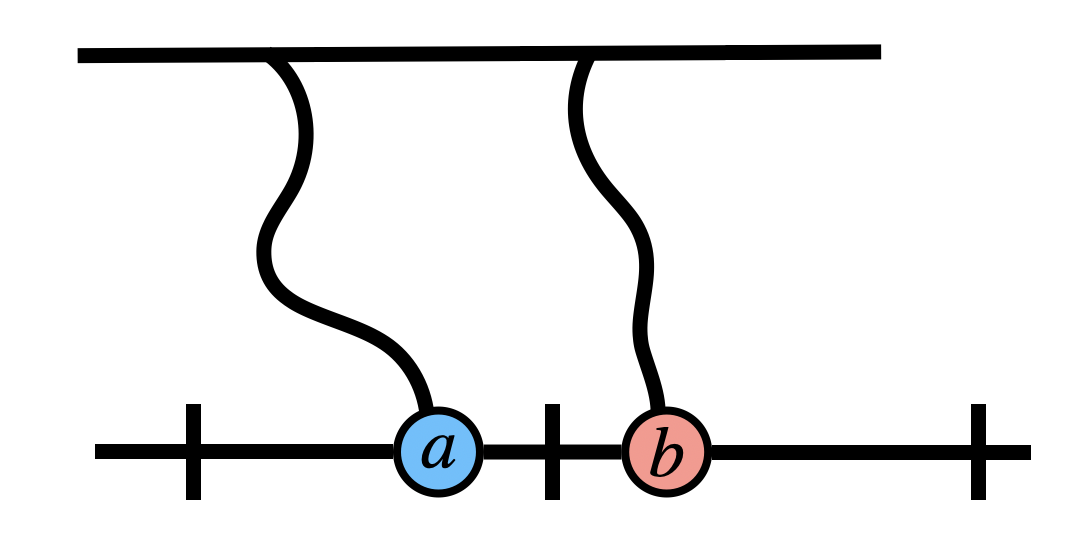}
        \caption{Shortcuts to pieces in \cite{hershkowitz2026planar}'s LC-ST algorithm.}
        \label{subfig:undirpieces1}
    \end{subfigure}%
    ~ 
    \begin{subfigure}[t]{0.5\textwidth}
        \centering
        \includegraphics[width=.5\linewidth]{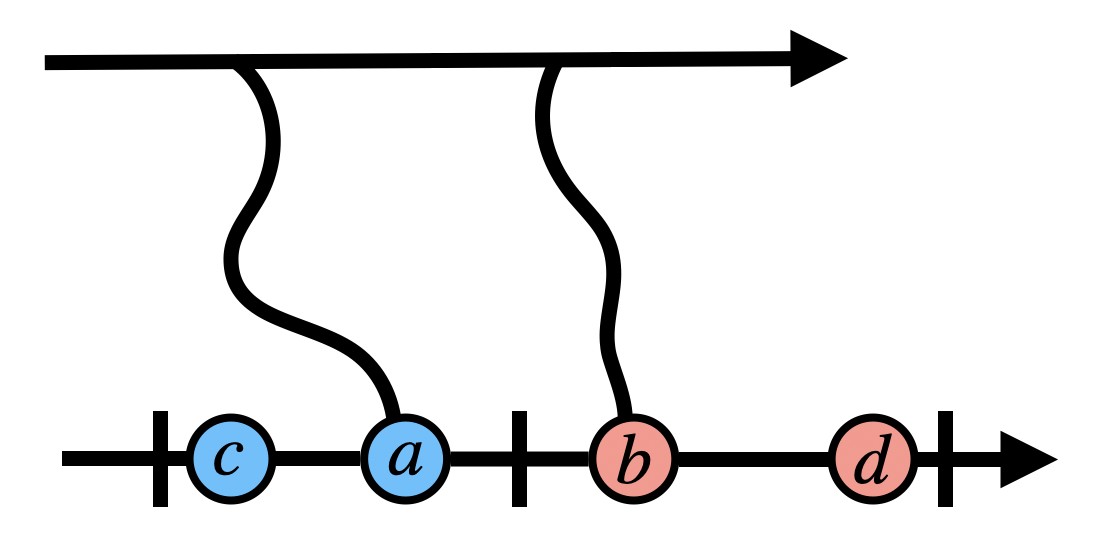}
        \caption{A failed directed analogue for pieces.}
                \label{subfig:undirpieces2}
    \end{subfigure}%
    \caption{The shortcuts to pieces in the algorithm of \citet{hershkowitz2026planar} for LC-ST and how it fails to generalize to LC-DST. \Cref{subfig:undirpieces1}: a horizontal parent and child separator path, and two pieces on the child separator whose shortcuts enter on vertices $a,b$. \Cref{subfig:undirpieces2}: if the separators are directed paths, then $a$ cannot reach vertices in its piece that lie before it on the path like $c$, but $b$ can still reach $d$.}
    \label{fig:undirpieces}
\end{figure}

\paragraph{Directed Challenges.} However, it is not clear how to extend the separator hierarchy and shortcuts framework of \citet{hershkowitz2026planar} to the directed setting or to obtain a cost of $ \approx O(\log ^ 2 k)$.
In more detail, the approach of \citet{hershkowitz2026planar} works as follows. For a given region in the hierarchy, one can break the associated length-constrained separator into $\approx \log k$ pieces of low-length and buy a shortcut from the root to each piece. This shortcut is a shortest length path from the parent separator to the given piece, restricted to paths with cost $O(\log k)\text{OPT}$. 
This necessarily costs a $O(\log^3k)$ factor in approximation, where one $\log k$ factor is from the height of the hierarchy, one is from the number of low-length pieces, and one is from the cost of the shortcuts. Moreover, obtaining near-exact length slack relies on the graph being undirected. This is because a shortcut path is allowed to reach a piece at any vertex. However, in directed graphs, this vertex may not be able to reach all other vertices of the piece.
It is not at all clear that this should extend to LC-DST. 
See also \Cref{fig:undirpieces}.

\begin{figure}[h]
    \centering
    \begin{subfigure}[t]{0.33\textwidth}
        \centering
        \includegraphics[width=.75\linewidth]{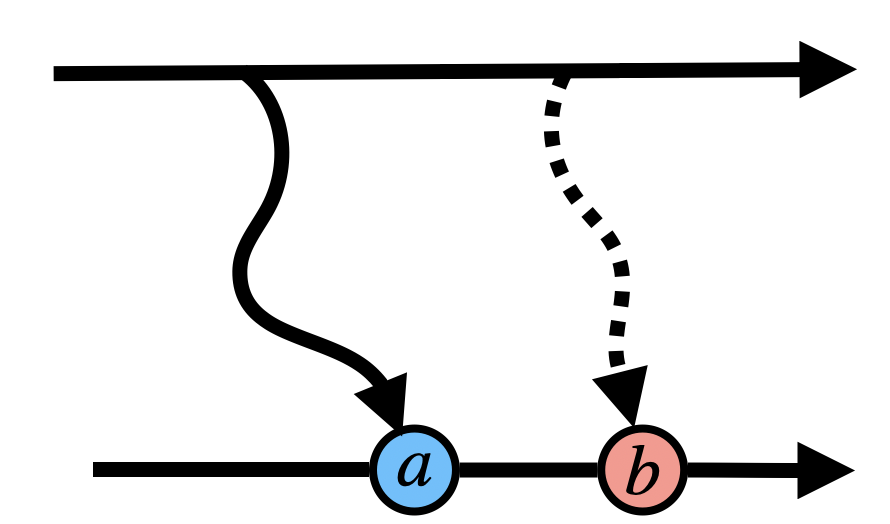}
        \caption{Candidate shortcut to $b$.}
        \label{subfig:purchaserule1}
    \end{subfigure}%
    ~ 
    \begin{subfigure}[t]{0.33\textwidth}
        \centering
        \includegraphics[width=.75\linewidth]{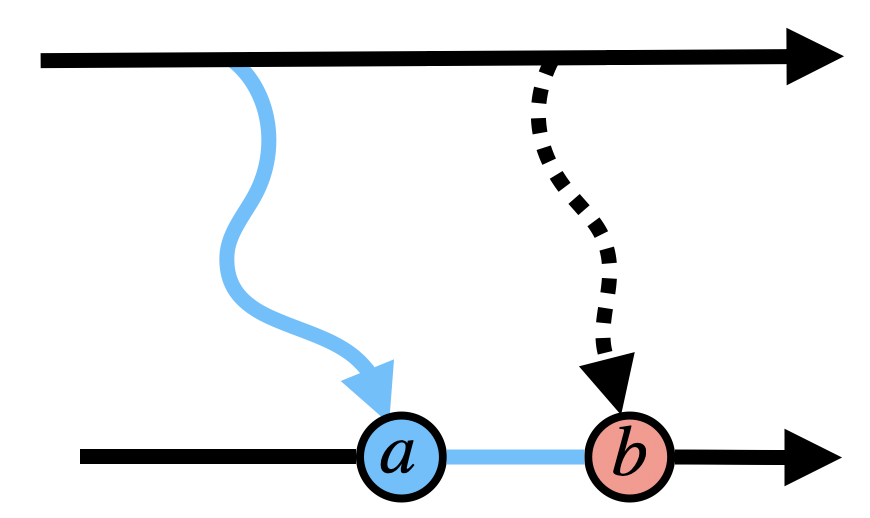}
        \caption{Path to $b$ using a previous shortcut.}
                \label{subfig:purchaserule2}
    \end{subfigure}%
    ~
    \begin{subfigure}[t]{.33\textwidth}
        \centering
        \includegraphics[width=.75\linewidth]{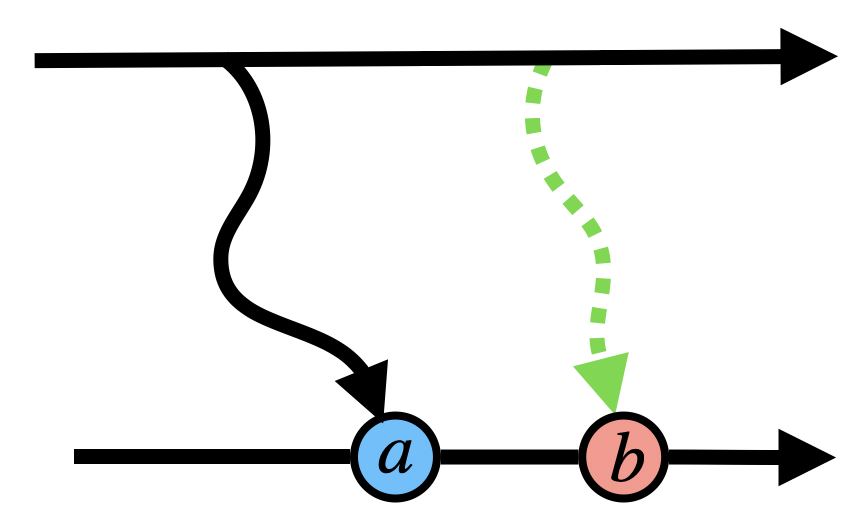}
        \caption{Path to $b$ using a new shortcut.}
        \label{subfig:purchaserule3}
    \end{subfigure}%
    \caption{The way we buy shortcuts. \Cref{subfig:purchaserule1}: horizontal parent and child separators, and a previously bought shortcut to $a$. The candidate shortcut from the parent separator to $b$ is in dotted black. \Cref{subfig:purchaserule2,subfig:purchaserule3}: two possible paths to reach $b$, one using the previously bought shortcut to $a$ and the subpath on the child separator from $a$ to $b$, and one using the candidate shortcut from the parent to $b$; we buy the candidate shortcut to $b$ if it is faster (under the lengths) for the solution to reach $b$ when taking the shortcut versus using the previous shortcut to $a$ and then the $a$-$b$ subpath on the separator.}
    \label{fig:purchaserule}
\end{figure}

\paragraph{Our Approach.} Our algorithm bypasses the need for fixed $O(\log k)$ pieces and cost bounds on shortcuts, by instead greedily choosing shortcuts according to a \textit{hierarchical mixture metric}. 
In particular, the algorithm of \citet{hershkowitz2026planar} defines the pieces first, and then purchases shortcuts for each piece. Here, we purchase shortcuts on-the-fly, where each purchase depends on our previous purchases. Each vertex on the separator will have a \textit{candidate shortcut} that we may or may not buy and add to our solution. Our purchases generate a partition of the separator into pieces that are non-uniform in length, each of which is a subpath going in a single direction. In particular, each shortcut generates a piece by entering the piece on its first vertex, so the shortcut can reach everything else in the piece. See also \Cref{fig:purchaserule}, where a second piece is generated if we buy the candidate shortcut.

The condition in which we buy a candidate shortcut is as follows. We scan the vertices of the child separator in the order they appear on the directed separator path. For each vertex $v$, its candidate shortcut is the shortest path under the hierarchical mixture metric from the parent separator to $v$. When scanning $v$, it is currently reachable via the closest previously bought shortcut and the path on the child separator. We buy its candidate shortcut if it is faster to reach $v$ via this shortcut instead. To recap, where the algorithm of \citet{hershkowitz2026planar} sets a budget for shortcut cost and optimizes over length, we instead optimize over a ``mixture'' of both cost and length and purchase shortcuts greedily. 

\paragraph{Analysis.}The algorithm's analysis uses a novel cost-to-length and length-to-cost charging framework that depends on the hierarchical mixture metric. In particular, we charge the cost of the shortcuts our solution buys to how much they affect the length of the solution when we buy them, and we charge the length of any root-terminal path in the final solution to the LP cost in each region of the separator hierarchy that it enters.  

\paragraph{Charging Cost to Length.} For the cost analysis, it suffices to bound the cost of the set of shortcuts we buy for a given region, since the cost of the separator hierarchy can be easily bounded to $\log^2k$ as done in earlier work \cite{hershkowitz2026planar,chekuri2026lcdsf}. Intuitively, each shortcut we buy should ``improve'' the length of the partial solution, otherwise buying it is pointless. We can exploit structure in these shortcuts, since these are shortest paths under the hierarchical mixture metric; namely, the cost of each shortcut we buy is proportional to its improvement to the solution's length. In particular, since we buy the shortcuts in directed order of the separator vertices, each purchase must improve the length from the root to the \textit{last} vertex on the separator, so all of the purchases together can only improve the length to the last vertex by at most $O(h)$. We charge each separator cost to these improvements in length. Since the total length improvement is at most $O(h)$, by appropriate scaling, we can bound the cost of shortcuts by $O(\log k)$ times the regional fractional cost.
Hence, this approach merges the $\log k$ factor from the fixed number of pieces into the $\log k$ factor for the cost of the shortcuts.

\paragraph{Charging Length to Cost.} To bound the length slack, for each terminal $t$, we define a \emph{canonical path} from the root to $t$. This is an $h$-length-bounded path which, (on average) for each region of the hierarchy, has cost proportional to the LP cost on that region.
For a given region $H$ and terminal $t$, let $P^*_t$ (resp.\ $P_t$) be the canonical path from the root to $t$ (resp.\ our solution's path from the root to $t$), and $P^*_{t,H}$ (resp.\ $P_{t,H}$) be the subpath of $P^*_t$ (resp.\ $P_t$) in $H$. We will compare the length of $P^*_{t,H}$ with the length of $P_{t,H}$. The solution's subpath $P_{t,H}$ consists of either a shortcut or a concatenation of a shortcut and a subpath of the separator. Because we are taking shortest paths with respect to the hierarchical mixture metric, we can argue that the increase in the length of $P_{t,H}$ as compared to $P^*_{t,H}$ is proportional to the cost of $P^*_{t,H}$. Since $P^*_t$ has length overall at most $h$, we get that the length of $P_t$ overall is at most $h$ plus a term proportional to the cost of $P^*_t$. Taking appropriate coefficients in our hierarchical mixture metric makes this factor at most $\epsilon h$.

\subsubsection{Length-Constrained Steiner Forest}

We build on the well-known \emph{junction tree} technique via new analysis that allows for $1+\epsilon$ length slack, and, more importantly, is LP-competitive.

\begin{figure}[t]
    \centering
\begingroup%
\pgfmathsetlengthmacro{\junctionxunit}{(\linewidth-40pt)/16}%
\definecolor{junctionred}{RGB}{205,55,45}%
\definecolor{junctiongreen}{RGB}{53,143,66}%
\definecolor{junctionblue}{RGB}{35,105,195}%
\definecolor{junctionpurple}{RGB}{133,81,175}%
\definecolor{junctionochre}{RGB}{193,128,20}%
\definecolor{junctionteal}{RGB}{0,142,156}%
\definecolor{junctionmagenta}{RGB}{192,62,133}%
\begin{tikzpicture}[
  x=\junctionxunit,y=1cm,
  line cap=round,line join=round,
  tree edge/.style={draw=black,line width=1.2pt},
  vertex/.style={inner sep=0pt,outer sep=0pt,minimum size=5pt},
  terminal label/.style={text height=1.5ex,text depth=.5ex},
  every node/.style={font=\normalfont\normalsize,inner sep=1pt}
]
  \coordinate (s1) at (.05,1.55);
  \coordinate (t4) at (1.75,1.35);
  \coordinate (a-upper) at (.95,.90);
  \coordinate (a-left) at (1.45,.05);
  \coordinate (s2) at (.10,-.85);
  \coordinate (v1) at (3.40,.30);
  \coordinate (a-right) at (5.45,.95);
  \coordinate (t1) at (6.20,1.60);
  \coordinate (s3) at (7.30,.15);
  \coordinate (a-bottom) at (4.40,-.65);
  \coordinate (s4) at (2.80,-1.40);
  \coordinate (a-lowerright) at (5.90,-1.05);
  \coordinate (t2) at (5.40,-1.55);
  \coordinate (t3) at (7.15,-1.25);

  \foreach \p/\q in {s1/a-upper,t4/a-upper,a-upper/a-left,s2/a-left,
                      a-left/v1,v1/a-right,a-right/t1,a-right/s3,
                      v1/a-bottom,a-bottom/s4,a-bottom/a-lowerright,
                      a-lowerright/t2,a-lowerright/t3}
    \draw[tree edge] (\p) -- (\q);

  \coordinate (s5) at (9.25,1.10);
  \coordinate (s6) at (8.85,-.95);
  \coordinate (b-left) at (10,.05);
  \coordinate (v2) at (10.60,.70);
  \coordinate (b-right) at (11.55,-.15);
  \coordinate (t5) at (12.70,.55);
  \coordinate (t6) at (11.90,-1.45);
  \foreach \p/\q in {s5/b-left,s6/b-left,b-left/v2,v2/b-right,
                      b-right/t5,b-right/t6}
    \draw[tree edge] (\p) -- (\q);

  \coordinate (s7) at (14,-.85);
  \coordinate (c-bend) at (14.65,-.05);
  \coordinate (v3) at (14.35,.70);
  \coordinate (t7) at (16,1.15);
  \foreach \p/\q in {s7/c-bend,c-bend/v3,v3/t7}
    \draw[tree edge] (\p) -- (\q);

  \foreach \p in {a-upper,a-left,v1,a-right,a-bottom,a-lowerright,
                   b-left,v2,b-right,c-bend,v3}
    \node[vertex,circle,fill=black] at (\p) {};
  \foreach \i/\col in {1/junctionred,2/junctiongreen,3/junctionblue,
                        4/junctionpurple,5/junctionochre,6/junctionteal,
                        7/junctionmagenta} {
    \node[vertex,rectangle,fill=\col] at (s\i) {};
    \node[vertex,rectangle,fill=\col] at (t\i) {};
  }

  \foreach \p/\lab/\col in {s1/s_1/junctionred,t4/t_4/junctionpurple,
                            t1/t_1/junctionred,s5/s_5/junctionochre,
                            t5/t_5/junctionochre,t7/t_7/junctionmagenta}
    \node[text=\col,anchor=base] at ([yshift=7pt]\p) {$\lab$};
  \foreach \p/\lab/\col in {s4/s_4/junctionpurple,t2/t_2/junctiongreen,
                            s7/s_7/junctionmagenta}
    \node[text=\col,anchor=base] at ([yshift=-12pt]\p) {$\lab$};
  \foreach \p/\lab/\col in {s2/s_2/junctiongreen,s6/s_6/junctionteal}
    \node[terminal label,text=\col,anchor=east]
      at ([xshift=-7pt]\p) {$\lab$};
  \foreach \p/\lab/\col in {s3/s_3/junctionblue,t3/t_3/junctionblue,
                            t6/t_6/junctionteal}
    \node[terminal label,text=\col,anchor=west]
      at ([xshift=7pt]\p) {$\lab$};

  \node[anchor=base] at ([yshift=7pt]v1) {$v_1$};
  \node[anchor=base] at ([yshift=7pt]v2) {$v_2$};
  \node[terminal label,anchor=east] at ([xshift=-7pt]v3) {$v_3$};
\end{tikzpicture}%
\endgroup%
    \caption{Three junction trees rooted at $v_1, v_2, v_3$ covering terminal pairs.}
    \label{fig:junction}
\end{figure}
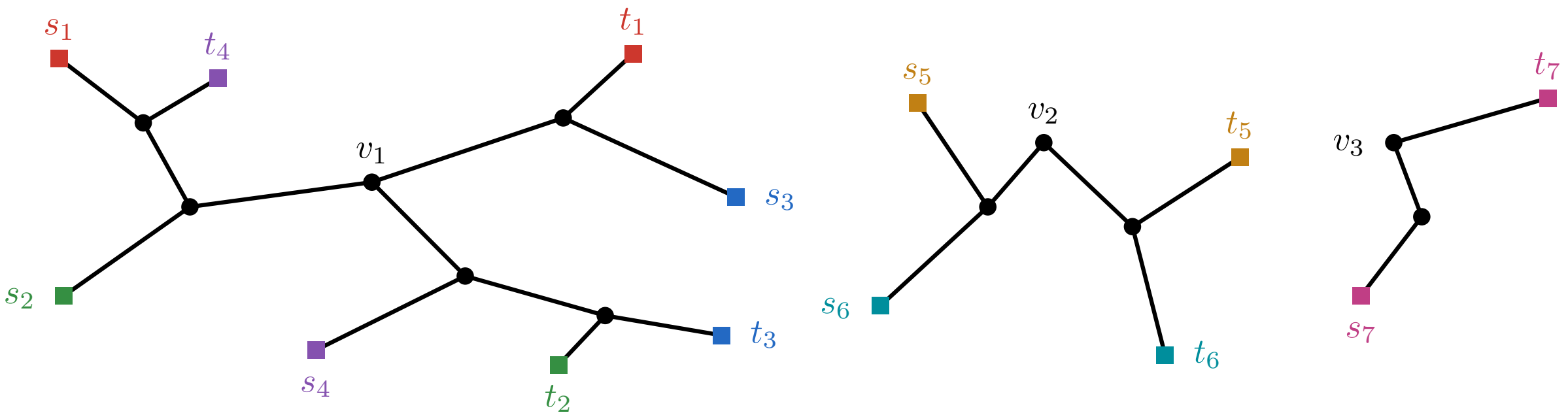

\paragraph{Junction Trees.} The junction tree technique was used by several earlier works in LC-SF (e.g. \cite{chekuri2010approximation,chekuri2026lcdsf}). It reduces a multicommodity problem to its single-source counterpart. A junction tree is a \emph{partial solution} (i.e. a subgraph that is feasible for a subset of terminal pairs) that connects pairs through a \emph{junction} vertex, see \Cref{fig:junction}. The goal is to repeatedly find junction trees of \emph{low density}, where the density of a partial solution is the ratio of its cost to the number of terminal pairs it connects. Finding an approximate minimum-density junction tree reduces to an LC-ST instance with the junction vertex as the root. Thus, the challenge in employing this approach typically lies in proving the existence of a low-density junction tree. Chekuri and Jain \cite{chekuri2026lcdsf} used the junction tree technique to show that an LP-competitive polylogarithmic approximation with $O(1)$ length slack for LC-ST yields the same for LC-SF. Combining this with \Cref{thm:LCDST} gives us a polylogarithmic approximation with $O(1)$ length slack for LC-SF. 

Our key technical contributions are making the approximation LP-competitive and reducing the length slack to $1 + \epsilon$, while also improving the logarithmic factors in the approximation ratio. We remark that obtaining LP-competitive bounds via the junction-tree approach has traditionally been challenging \cite{chekuri2010approximation}, even in planar graphs \cite{chekuri2025dsf}. This is because the existence of a low-density junction tree is typically proved by decomposing an integral optimum solution into junction trees, and many prior works have failed to construct such decompositions with respect to fractional solutions. 

\begin{figure}[t]
    \centering
\begingroup%
\pgfmathsetlengthmacro{\approachxunit}{(\linewidth-12pt)/16}%
\definecolor{pairone}{RGB}{205,55,45}%
\definecolor{pairtwo}{RGB}{53,143,66}%
\definecolor{pairthree}{RGB}{35,105,195}%
\begin{tikzpicture}[
  x=\approachxunit,y=1cm,
  line cap=round,line join=round,
  flow/.style={line width=1.2pt},
  vertex/.style={inner sep=0pt,outer sep=0pt,minimum size=5pt},
  every node/.style={font=\normalfont\normalsize,inner sep=1pt}
]
  \draw[line width=1.7pt] (0,0) -- (16,0);
  \foreach \k in {1,...,19} {
    \pgfmathsetmacro{\x}{.8*\k}
    \draw[line width=.85pt]
      ([yshift=-4pt]\x,0) -- ([yshift=4pt]\x,0);
  }
  \node[vertex,circle,fill=black] at (0,0) {};
  \node[vertex,circle,fill=black] at (16,0) {};
  \node[anchor=base] at ([yshift=7pt]0,0) {$Q$};
  \draw[decorate,decoration={brace,mirror,amplitude=2.5pt},
        line width=.75pt]
    ([yshift=-5.5pt].8,0) -- ([yshift=-5.5pt]1.6,0)
    node[midway,below=4pt] {$l(C)\leq\epsilon h$};

  \coordinate (s1) at (3.80,-1.45);
  \coordinate (t1) at (3.55,1.55);
  \draw[flow,pairone] (s1)
    .. controls (3.35,-1.02) and (3.02,-.45) .. (3.05,0)
    .. controls (3.10,.55) and (3.46,1.18) .. (t1);
  \draw[flow,pairone] (s1)
    .. controls (3.52,-1.00) and (3.30,-.48) .. (3.34,0)
    .. controls (3.34,.60) and (3.60,1.20) .. (t1);
  \draw[flow,pairone] (s1)
    .. controls (3.72,-.90) and (3.52,-.42) .. (3.58,0)
    -- (3.82,0)
    .. controls (3.82,.52) and (3.74,1.14) .. (t1);
  \draw[flow,pairone] (s1)
    .. controls (4.05,-1.04) and (4.18,-.48) .. (4.16,0)
    .. controls (4.19,.57) and (3.98,1.16) .. (t1);

  \coordinate (s3) at (4.80,-1.45);
  \coordinate (t3) at (5.70,-1.45);
  \draw[flow,pairthree] (s3)
    .. controls (4.52,-1.05) and (4.34,-.55) .. (4.42,0)
    .. controls (4.37,.68) and (4.50,1.02) .. (5.02,1.02)
    .. controls (5.50,1.02) and (5.80,.65) .. (5.78,0)
    .. controls (5.76,-.54) and (5.75,-1.10) .. (t3);
  \draw[flow,pairthree] (s3)
    .. controls (4.85,-1.00) and (4.69,-.36) .. (4.98,0)
    .. controls (5.15,-.40) and (5.28,-1.15) .. (t3);

  \coordinate (s2) at (12.30,-1.45);
  \coordinate (t2) at (12.50,1.55);
  \draw[flow,pairtwo] (s2)
    .. controls (11.90,-1.00) and (11.70,-.45) .. (11.74,0)
    .. controls (11.74,.65) and (12.18,1.20) .. (t2);
  \draw[flow,pairtwo] (s2)
    .. controls (12.12,-1.00) and (12.10,-.45) .. (12.14,0)
    .. controls (12.14,.65) and (12.32,1.20) .. (t2);
  \draw[flow,pairtwo] (s2)
    .. controls (12.48,-1.00) and (12.51,-.45) .. (12.54,0)
    .. controls (12.56,.65) and (12.47,1.20) .. (t2);
  \draw[flow,pairtwo] (s2)
    .. controls (12.83,-1.00) and (12.98,-.45) .. (12.94,0)
    .. controls (12.92,.65) and (12.72,1.20) .. (t2);
  \draw[flow,pairtwo] (s2)
    .. controls (13.22,-1.00) and (13.84,-.45) .. (13.78,0)
    .. controls (13.74,.65) and (13.05,1.20) .. (t2);

  \foreach \x in {3.05,3.34,3.58,3.82,4.16}
    \node[vertex,circle,fill=pairone] at (\x,0) {};
  \foreach \x in {4.42,4.98,5.78}
    \node[vertex,circle,fill=pairthree] at (\x,0) {};
  \foreach \x in {11.74,12.14,12.54,12.94,13.78}
    \node[vertex,circle,fill=pairtwo] at (\x,0) {};

  \foreach \p/\col in {s1/pairone,t1/pairone,s3/pairthree,t3/pairthree,
                       s2/pairtwo,t2/pairtwo}
    \node[vertex,rectangle,fill=\col] at (\p) {};
  \foreach \p/\col/\lab in {t1/pairone/t_1,t2/pairtwo/t_2}
    \node[text=\col,anchor=base] at ([yshift=7pt]\p) {$\lab$};
  \foreach \p/\col/\lab in {s1/pairone/s_1,s3/pairthree/s_3,
                            t3/pairthree/t_3,s2/pairtwo/s_2}
    \node[text=\col,anchor=base] at ([yshift=-12pt]\p) {$\lab$};
\end{tikzpicture}%
\endgroup%
    \caption{Example of flow paths intersecting separator $Q$, divided into chunks $C$. Terminals are represented by squares, all other nodes by circles. While $Q$ may be long and expensive, only a small number of chunks are used by each terminal pair.}
    \label{fig:overview_lcsf}
\end{figure}
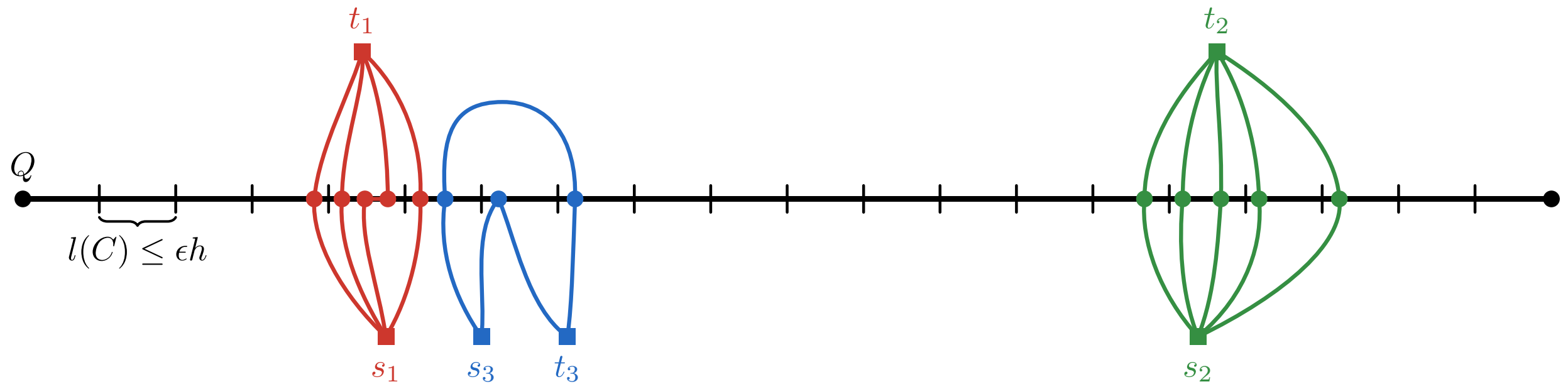

\paragraph{Our Approach.} We overcome this barrier and construct \emph{candidate partial solutions} that, on average, have low density with respect to the LP optimum. We then greedily choose min-density candidates until all pairs are covered. To construct the candidate solutions, we use the length-constrained separators. In single-source problems (LC-ST and LC-DST), the given root has short, low-cost paths to each terminal, allowing one to bound the cost of separator paths. Since LC-SF is a multicommodity problem, there is no choice of root with such a guarantee. Therefore, we construct the separator from an arbitrary root $r \in V$, but do not buy the separator paths. Instead, we show that for any separator path $Q$ and each terminal pair $(s_i, t_i)$, all $s_i$-$t_i$ flow given by the LP intersects $Q$ only on a small subpath; a similar argument was used by \citet{iglesias2018plane} on a related problem.\footnote{This claim only holds if all flow paths have roughly the same cost. We handle this by bucketing paths by cost and treating each bucket separately.} Thus, we can divide $Q$ into \emph{chunks} of length approximately $\epsilon h$, such that for each terminal pair, $s_i$-$t_i$ flow only passes through a small number of chunks -- see \Cref{fig:overview_lcsf}. These chunks, when contracted, serve as junctions from which to construct junction trees. 
\begin{itemize}
    \item \textbf{Cost Bound.} To bound the cost of the candidate partial solutions, we use a similar observation as above: for any separator path $Q$ and each edge $e \in E$, all LP flow paths containing $e$ intersect $Q$ in a small subpath. Therefore, we can charge the cost of each contracted chunk to the corresponding LP flow covering it without too much overlap. To improve logarithmic factors in the approximation ratio, our proof uses a more intricate weighted charging argument -- details are deferred to \Cref{sec:lcsf}.
    \item \textbf{Length Bound.} The junction tree approach typically loses at least a factor $2$ in the length slack, as $s_i$ and $t_i$ are each granted a budget of $h$ to reach the junction. Instead, we impose non-uniform length bounds $a_i$ for $s_i$ and $h - a_i$ for $t_i$. We choose $a_i$ to be the flow-weighted median distance from $s_i$ to the junction, ensuring that each endpoint retains at least half the flow to the junction with its new length constraint. This, along with the fact that each chunk length is at most $\epsilon h$, yields the desired $1+ \epsilon$ length slack.
\end{itemize}
\noindent
Once we have constructed all candidate solutions on the separator paths from $r$, we remove these separators from $G$ and recurse on remaining connected components, until all terminal pairs are separated.


\subsection{Related Work} 
\paragraph{Length-Constrained Network Design.}
Length-constrained network design was first introduced by \citet{ba92}, and has since been well-studied in approximation algorithms.
Problems in this space typically require bicriteria approximation algorithms to obtain reasonable guarantees. This is not surprising, as even the basic building block of computing length-constrained distances (also known as \textit{restricted shortest paths} in the literature) is NP-hard \cite{garey2002computers}, but can be efficiently approximated within $1+\epsilon$ \cite{hassin1992approximation}.\footnote{Restricted shortest paths is solvable in polynomial time when the lengths are polynomially bounded via Bellman-Ford.} LC-ST is NP-hard even when the input is restricted to planar graphs \cite{hershkowitz2026planar,friggstad2023dst}. Even worse, LC-ST has a bicriteria hardness of approximation in general graphs: conditioned on a popular conjecture in complexity theory, there is no $\log^{2-\epsilon}n$ approximation for LC-ST with length slack strictly less than $2$ \cite{hershkowitz2026planar}. A slight modification to that reduction immediately gives the same statement for LC-ST but with length slack strictly less than $3$, and for LC-DST but with any length slack. 

On the positive side, LC-ST admits a $O(\log k)$ approximation with $O(\log k)$ length slack via a simple clustering approach by \citet{marathe1998bicriteria}, and LC-SF admits a $\polylog(k)$ approximation with $\polylog(k)$ length slack via connections to buy-at-bulk network design \cite{chekuri2010approximation,chekuri2024hcbab} and hop-constrained tree embeddings \cite{haeupler2021treeemb,filtser2022hop}---both results for LC-ST and LC-SF can be made LP-competitive \cite{khani2016improved,chekuri2024hcbab,ghaffari2021hop}. 

True approximation algorithms for LC-ST are known essentially by its equivalence for DST, which has had more success in the quasipolynomial time regime.  In particular, if the length slack $\beta$ is $1$ then we say the algorithm is \textit{length-exact}. The best known length-exact polynomial time approximation is a $O(k^{\epsilon}/\epsilon^3)$ approximation with running time dependent on $\epsilon$ by \citet{Zelikovsky97,charikar1999dst}. Trading off with $\epsilon$, this gives a quasipolynomial time $O(\log^3k)$ approximation. This was improved by \citet{grandoni2019log2} to a tight (under the projection games conjecture) quasipolynomial time $O(\log^2k/\log\log k)$ approximation using an LP hierarchy approach. This result was later replicated a recursive greedy algorithm that also works for LC-DST by \citet{ghuge2022quasi}.

\paragraph{Planar Graphs.}
\citet{demaine2009node} first considered planar DST by providing a non-crossing flow relaxation that could be rounded with $O(1)$ loss, but it is not known how to solve the relaxation. \citet{friggstad2023qbmfdst} later gave an $O(1)$ approximation for \textit{quasi-bipartite} minor-free instances of DST with a primal-dual approach, followed by an $O(\log k)$ approximation for planar DST with a different but simple approach using planar shortest path separators \cite{friggstad2023dst}. This bypassed the $\Omega(\log^2k)$ hardness of DST in general graphs \cite{halperin2003polylogarithmic}, showing that DST is easier in planar graphs. This was further extended to an LP-competitive $O(\log^2k)$ approximation for planar DST by \citet{chekuri2024dpst}, who also gave polylogarithmic approximations for a variety of single-source planar network design problems, including directed polymatroid Steiner tree. Using the LP-competitive planar DST algorithm as a subroutine, \citet{chekuri2025dsf} gave an $O(\log^6k)$ approximation for planar DSF, also bypassing its hardness of approximation in general graphs.

The success for these problems in planar graphs was then emulated for the length-constrained counterparts. The planar shortest path separator approach alone is not directly useful for length-constrained problems, since a shortest path under the costs could be long under the lengths, and vice versa. \citet{hershkowitz2026planar} observed that this could be remedied using the \textit{mixture metric}, which had previously been used in related problems \cite{meyerson2000cost,haeupler2021treeemb}. Combining planar separators with the mixture metric yields a notion of \textit{length-constrained separators}. However, following the approach of \citet{friggstad2023dst} with these stronger objects still only gives an $O(\log k)$-approximation with $O(\log k)$ length slack, which is already attainable in general graphs. With more work (see \Cref{sec:tech_overview}), \citet{hershkowitz2026planar} obtain an $O(\log^{1+\epsilon}k)$ approximation (and LP-competitive $O(\log^3k/\epsilon)$ approximation) with $1+\epsilon$ length slack, bypassing the aforementioned bicriteria hardness of approximation for LC-ST in general graphs. Using \textit{directed} length-constrained separators, \citet{chekuri2026lcdsf} obtained an $O(\log k)$ approximation (and LP-competitive $O(\log^2 k)$ approximation) with $O(\log k)$ length slack for planar LC-DST. They also carefully modify the planar DSF algorithm to get a $O(\log^6k)$ approximation with $O(\log k)$ length slack for planar LC-DSF.

\paragraph{Other Related Problems.}
The cost-distance and buy-at-bulk network design problems also involve two edge-weight functions and have been well-studied (see e.g. \cite{batb97,awerbuch1997buy,meyerson2000cost,chekuri2001costdist,bab_random_greedy,chekuri2010approximation}). In fact, algorithms for buy-at-bulk network design generally immediately imply algorithms for length-constrained network design and vice-versa; see e.g.\ \cite{chekuri2024hcbab} for more details. 
The field of length-constrained graph algorithms also extends far beyond network design, with many recent breakthroughs in length-constrained flows and routing schemes based on the developing theory of length-constrained expanders and expander decompositions; see e.g.\ \cite{haeupler2024new} for an overview. 

\section{Preliminaries}\label{sec:prelim}
In this section, we provide the necessary notations and existing results that we will use. The following two sections are dedicated to our planar LC-DST algorithm (\Cref{sec:lcdst}) and planar LC-SF algorithm (\Cref{sec:lcsf}). 
\subsection{Notation}
\paragraph{Graphs.}
We consider both undirected and directed graphs, and write $G=(V,E)$ in both cases, with $n:=|V|$. For a $S\subseteq V$ we let $G[S]$ be the induced subgraph on $S$ in $G$, and let $\delta(S):=\{(u,v)\in E:u\in S,v\not\in S\}$ when $G$ is undirected and $\delta^+(S):=\{(u,v)\in E:u\in S,v\not\in S\},\delta^-(S):=\{(u,v)\in E:u\not\in S,v\in S\}$ when $G$ is directed.   
An \emph{out-arborescence rooted at $r$} is a directed tree in which every vertex has a unique directed path from $r$. A graph is \emph{planar} if it can be drawn in the plane without crossing edges. For a path $P \subseteq G$ with $u,v \in V(P)$ we let $P[u,v]$ denote the subpath of $P$ from $u$ to $v$. A weakly connected component of a directed graph $G$ is a connected component of the underlying undirected graph of $G$. For brevity we call these connected components. 

\paragraph{Costs and Lengths.}
Every edge $e\in E$ has a \emph{cost} $c(e)\in\mathbb{Z}_{\ge0}$ and a \emph{length} $l(e)\in\mathbb{Z}_{\ge0}$. For $F\subseteq E$ we write
$c(F):=\sum_{e\in F}c(e), l(F):=\sum_{e\in F}l(e)$
and for a subgraph $H$ we write $c(H):=c(E(H))$ and $l(H):=l(E(H))$. 
 
\paragraph{Distances.}
For a subgraph $H$ and $u,v\in V(H)$, let $d_H(u,v)$ be the length of a shortest $u,v$ path in $H$ with respect to $l$, with $d_H(u,v)=\infty$ if no such path exists. We write $d$ for $d_G$. 
 
For a length bound $h\geq1$, let $\mathcal{P}_H^{(h)}(u,v)$ be the set of $u,v$ paths $P$ in $H$ with $l(P)\le h$. The \emph{$h$-length-constrained distance} from $u$ to $v$ in $H$ is the minimum cost of such a path,
\[
d_H^{(h)}(u,v):=\min\big\{\,c(P)\;:\;P\in\mathcal{P}_H^{(h)}(u,v)\,\big\},
\]
with $d_H^{(h)}(u,v)=\infty$ if $\mathcal{P}_H^{(h)}(u,v)=\emptyset$. We write $d^{(h)}$ for $d^{(h)}_G$. Thus $d_H^{(h)}(u,v)<\infty$ if and only if $d_H(u,v)\le h$.
The \emph{$h$-length-constrained radius} of $H$ from a vertex $r$ is
\[
D_H^{(h)}(r):=\max_{v\in V(H)}d_H^{(h)}(r,v).
\]
 
\paragraph{Problem Definitions.}
All four problems we consider take as input a graph $G=(V,E)$ with costs $c$ and lengths $l$ and an integer length bound $h>0$. They differ in whether $G$ is directed and in the connectivity requirement. Throughout, $k$ denotes the number of terminals or demand pairs.
 
In \textbf{length-constrained Steiner tree (LC-ST)}, $G$ is undirected and we are further given a root $r\in V$, terminals $S\subseteq V\setminus\{r\},|S|=k$. The goal is a minimum cost subgraph $T\subseteq G$ with $d_T(r,t)\le h$ for every $t\in S$. 
In \textbf{length-constrained Steiner forest (LC-SF)}, $G$ is undirected, and we are further given demand pairs $(s_1,t_1),\dots,(s_k,t_k)$ of vertices. The goal is a set $F\subseteq E$ that satisfies $d_F(s_i,t_i)\le h$ for every $i\in\{1,\dots,k\}$ and minimizes $c(F)$. The \emph{uniform} case is $h_1=\dots=h_k=h$.
Unlike in the unconstrained problem, an optimal LC-SF solution need not be a forest. For example, on a cycle with unit lengths whose consecutive vertices form demand pairs with bound $1$, every edge of the cycle is required.
The directed analogues LC-DST/LC-DSF are the same, but $G$ is directed and the length requirement must be satisfied with respect to directed paths.

We write $\mathrm{OPT}_h$ for the minimum cost of a feasible solution for whichever problem we are working on, and use $\operatorname{OPT}$ for short when $h$ is clear. A solution to any of the problems is $h$-length if it is feasible and $\beta h$-length if it violates the length constraint $h$ by a multiplicative $\beta$ factor. For $\alpha,\beta\ge1$, an \emph{$\alpha$ approximation} with $\beta$ length slack returns a $\beta h$-length solution of cost at most $\alpha\cdot\mathrm{OPT}_h$. If the cost is instead bounded by $\alpha$ times the optimal value of a linear programming relaxation of the problem, the guarantee is \emph{LP-competitive}.

\subsection{Length-Constrained Separators} 
In this subsection, we provide necessary background on length-constrained planar separators, which will be used repeatedly in our algorithms.
\paragraph{The Mixture Metric and Length-Constrained Separators.}
We define the general form of the mixture metric that has been used in earlier work \cite{meyerson2000cost,chekuri2026lcdsf}. Roughly, the mixture metric is a metric that balances the two edge-weight functions.

Fix $h,\gamma>0$. The \emph{mixture weight} of an edge $e$ is
\[
w_{mix}(e):=c(e)+\frac{\gamma}{h}\,l(e).
\]
For a subgraph $H$ and $u,v\in V(H)$, let $d^{mix}_H(u,v)$ be the minimum mixture weight of a $u,v$ path in $H$, and write $d^{mix}$ for $d^{mix}_G$. We use the following standard properties.
 
\begin{lemma}[Mixture metric; {\cite{hershkowitz2026planar,chekuri2026lcdsf}}]\label{lem:mixture}
    $d^{mix}$ forms a metric. Furthermore, for every path $P$ in $G$:
    \begin{enumerate}
        \item \textbf{Mixture Lower Bound}: $c(P)\le w_{mix}(P)$ and $l(P)\le\frac{h}{\gamma}\,w_{mix}(P)$;
        \item \textbf{Mixture Upper Bound}: If $l(P)\le h$, then $w_{mix}(P)\le c(P)+\gamma$.
    \end{enumerate}
\end{lemma}
\noindent 
Since the mixture distances are a metric whose shortest paths exhibit these desired bounds, it is natural to combine them with shortest path separators.
\begin{lemma}[Separators; \cite{lipton1979separator,thorup2004compact}]\label{lem:separator}
    Let $G$ be a connected (directed or undirected) planar graph with vertex weights $w:V\to\mathbb{R}_{\ge0}$, and let $T$ be a spanning tree of $G$ rooted at $r$. There is a polynomial-time algorithm that finds $u_1,u_2,u_3\in V$ such that every connected component of $G-V\big(T[r,u_1]\cup T[r,u_2]\cup T[r,u_3]\big)$ has weight at most $w(V)/2$.
\end{lemma}
\noindent
Combining the previous two lemmas gives the following:
\begin{lemma}[Length-constrained separators; \cite{hershkowitz2026planar,chekuri2026lcdsf}]\label{lem:lc-separator}
    Let $G$ be a (directed or undirected) planar graph with costs $c$, lengths $l$, a root $r$ and terminals $S\subseteq V$. Fix $h\ge1$, $\gamma>0$. In polynomial time one can compute paths $P_1,P_2,P_3$ in $G$, each starting at $r$ (and directed away from $r$ if $G$ is directed), with the following properties. Let $P:=P_1\cup P_2\cup P_3$.
    \begin{enumerate}
        \item \textbf{Balanced Separator}: Every connected component of $G-P$ has at most $|S|/2$ terminals. 
        \item \textbf{Length-Constrained}: For every $j\in\{1,2,3\}$ we have $P_j$ is a shortest path according to $w_{mix}$. Moreover for every vertex $v$ on $P_j$, we have $c(P_j[r,v])\leq O(\gamma)$ and $l(P_j[r,v])\le O(h)$.
    \end{enumerate}
\end{lemma}

\subsection{Linear Programming Relaxations}
Our algorithms will be approximate with respect to the following LP's for LC-DST and LC-SF. We can think of the relaxation as setting capacities $x_e$'s on the edges such that each demand has a unit flow over its paths. 
\begin{equation}\tag{LC-DST-LP}\label{lp:lcdst}
\begin{alignedat}{2}
\min\quad & \sum_{e\in E}c(e)\,x_e \\
\text{s.t.}\quad & \sum_{P\in\mathcal{P}^{(h)}(r,t)}f_P\ge1 &\qquad& \forall\,t\in S,\\
& \sum_{P\in\mathcal{P}^{(h)}(r,t):\,e\in P}f_P\le x_e && \forall\,t\in S,\ e\in E,\\
& x\ge0,\ f\ge0. &&
\end{alignedat}
\end{equation}
 \noindent
For LC-SF with demand pairs $(s_i,t_i)$ and length bound $h$, the relaxation is
\begin{equation}\tag{LC-SF-LP}\label{lp:lcsf}
\begin{alignedat}{2}
\min\quad & \sum_{e\in E}c(e)\,x_e \\
\text{s.t.}\quad & \sum_{P\in\mathcal{P}^{(h)}(s_i,t_i)}f_P\ge1 &\qquad& \forall\,i\in[k],\\
& \sum_{P\in\mathcal{P}^{(h)}(s_i,t_i):\,e\in P}f_P\le x_e && \forall\,i\in[k],\ e\in E,\\
& x\ge0,\ f\ge0. &&
\end{alignedat}
\end{equation}
The undirected LC-ST-LP and directed LC-DSF-LP counterparts have analogous relaxations.
 
Each relaxation has value at most $\mathrm{OPT}$ for its respective problems. They have exponentially many variables, but can be solved for approximately up to a $1+\epsilon$ factor efficiently using the ellipsoid method and restricted shortest paths as the separation oracle on their respective duals. The LP-competitive guarantees for LC-ST, LC-DST and LC-SF are with respect to \eqref{lp:lcdst} (or (LC-ST-LP)) and \eqref{lp:lcsf}.

\section{LP-Competitive Length-Constrained Directed Steiner Tree}\label{sec:lcdst}
In this section, we present our algorithm for LC-DST and prove the following theorem.
\thmLCDST*

\subsection{Algorithm}
Our algorithm rounds a fractional solution $(x,f)$ to \ref{lp:lcdst} (that without loss of generality can be assumed to have no edge $e$ with $x_e = 0$) to an integral solution that reaches every terminal from $r$ by a length $(1 + \epsilon)h$ path. The first step in doing so is constructing a planar separator hierarchy, defined below. We recall that a region of a graph is any edge-induced subgraph of it.
\begin{definition}[Separator Hierarchy]
\label{def:lcdst_separator_hierarchy}
Given a planar graph $G$ and a root $r\in V(G)$, a \textbf{separator hierarchy} of $G$ is a rooted tree $\mathcal{T}$ whose nodes correspond to regions of $G$ such that the children of a node $v$ correspond to the regions obtained by computing a planar separator of $v$'s region. Hence each node of $\mathcal{T}$ is also associated with a planar separator of its region. 
\end{definition}
\noindent
We provide some terminology we will use repeatedly. Given a node $v$ with parent node $u$ in $\mathcal{T}$, we call their respective regions (resp.\ associated separators) child/parent regions (resp.\ separators). The ancestors of a region $H$ are the regions corresponding to its ancestor nodes in the hierarchy. The ancestor subgraph $A_H$ of region $H$ is the union of $r$ and the separator paths corresponding to $H$'s ancestor nodes. We call vertices of $A_H$ \emph{ancestor vertices}. Let $I_H = V(H) \setminus V(A_H)$ be the $\textit{interior}$ and let $E_H = E(H)$; we denote the terminals $t \in S \cap I_H$ as the interior terminals of $H$. Initially in our recursion, $H = G$ and $A_H = \{r\}$.

We denote the LP cost restricted to a region $H$ as
\begin{align*}
\lpvalue & := \sum_{e \in E_H}c(e)x_e.
\end{align*}
Computing a hierarchy requires recursively computing shortest path separators; we compute these shortest paths under the following metric.
\begin{definition}[Hierarchical Mixture Metric]\label{defn:hierarchymix}
For a region $H$ with $\lpvalue > 0$, define the edge weights
\begin{align*}
w_H(e) & := \frac{l(e)}{\epsilon h} + \frac{c(e)}{\lpvalue\numlevels}.
\end{align*}
For a path $P$, let $w_H(P) = \sum_{e \in P} w_{H}(e)$. We define the \textbf{hierarchical mixture metric} to be the shortest path distance under these mixture weights.
\end{definition}
\noindent
After computing the hierarchy under the hierarchical mixture metric, we buy a set of ``shortcuts'' between every pair of separators corresponding to parent-child nodes in the hierarchy. Specifically, each vertex on the child separator is associated with some \textit{candidate shortcut}, a shortest path under the hierarchical mixture metric of its region. We will not necessarily buy all of the candidate shortcuts for a given (child) separator.

We next describe in more detail the intuition above.

\paragraph{Step 1: Constructing the Separator Hierarchy.}
Having defined our hierarchical mixture metric, we now use it to construct our separator hierarchy. Starting at the root of our separator hierarchy, we first prune vertices that are far according to our hierarchical mixture metric. We then recursively take shortest path planar separators also using our hierarchical mixture metric. We formalize these two steps below.


To obtain short, low-cost separators in each of our recursive calls, we first restrict the region $H$ for each call to only vertices close to its ancestor subgraph $A_H$ under our hierarchical mixture metric. If $\lpvalue = 0$, we do not buy a separator in this call and stop the recursion in $H$. For $\lpvalue > 0$, we form an auxiliary graph by temporarily adding the edges of the ancestor subgraph $A_H$ and contracting $A_H$ into the root $r$ as a single vertex, called the \emph{contracted root}. Using this contracted root, we define the pruned region of $H$ as follows.
\begin{definition}[Pruned Region]
\label{def:lcdst_pruning}
For $\lpvalue > 0$, we define $\operatorname{prune}(H)$ to be the subgraph of $H$ induced by all its ancestor vertices and interior vertices at distance at most $1 + 1/\epsilon$ from the contracted root under $w_H$, where this distance is measured in the graph obtained by contracting $A_H$ in $H \cup A_H$.
\end{definition}
\noindent
We call $\operatorname{prune}(H)$ the \emph{pruned region} and $1 + 1/\epsilon$ the \emph{pruning radius}. We then replace $H$ by $\operatorname{prune}(H)$ and continue to work with the pruned region, leaving $A_H, I_H, E_H$, and $\lpvalue$ unchanged. This auxiliary graph is planar, because the ancestor subgraph $A_H$ is connected in the underlying undirected graph of $G$, and thus the contraction preserves planarity.

After pruning $H$, we next recursively separate the terminals in its interior using shortest path planar separators under our hierarchical mixture metric. To do so, we apply \Cref{lem:separator} with weight 1 on interior terminals (see \Cref{def:lcdst_separator_hierarchy}) and 0 elsewhere. This gives us a planar separator consisting of at most three shortest paths whose removal forms connected components each with at most half the interior terminals. We undo the contraction and add its corresponding paths $\mathcal{Q} := Q_{1} \cup Q_{2} \cup Q_{3}$ to our solution $T$. We then recurse on each connected component of $H \setminus (A_{H} \cup \mathcal{Q})$ containing a terminal. \Cref{alg:lcdst_buyseparators} describes the pseudocode for recursively buying these separators.
We set the costs of the separator edges we bought to 0 for the remainder of the algorithm; $w_H$ now uses these modified costs, whereas $\lpvalue$\textit{ remains its original value}.

\begin{algorithm}[H]
    \caption{BuySeparators$(T, H,A_H)$}    
    \begin{algorithmic}[1]
        \label{alg:lcdst_buyseparators}
        \setlength{\itemsep}{1pt}
        \State Set $I_H,E_H,\lpvalue$ using \Cref{def:lcdst_separator_hierarchy}
        \State If $I_H \cap S = \emptyset$ or $\lpvalue = 0$, \Return 
        \State $H \gets \operatorname{prune}(H)$ using \Cref{def:lcdst_pruning}
        \State Let $\mathcal{Q} = Q_1 \cup Q_2 \cup Q_3$ be given by \Cref{lem:separator} 
        \State $T \gets T \cup \mathcal{Q}$
        \For{each connected component $C_i$ of $H \setminus (A_H \cup \mathcal{Q})$ containing a terminal}
            \State $\textsc{BuySeparators}(T,C_i \cup A_H \cup \mathcal{Q},A_H \cup \mathcal{Q})$
        \EndFor
    \end{algorithmic}
\end{algorithm}

\paragraph{Step 2: Finding Candidate Shortcuts.} We now buy shortcuts between the separators of our separator hierarchy. We consider regions in parent-before-child order. Within each region, we consider its separator paths one at a time, and consider the vertices $u$ on each separator path $Q = (q_{0},\ldots, z)$ in the order of following its outgoing edges from $q_{0}$ to $z$.

For each vertex $u$ on a separator $Q$ in region $H$ of our separator hierarchy (with $\lpvalue > 0$), we define a candidate shortcut path $P_u$ that we will consider purchasing. We define $P_u$ to be the shortest path from some ancestor vertex $a_u$ to $u$ under the hierarchical mixture metric, while also accounting for the length to reach $a_u$ from the root $r$, as follows.
\begin{definition}[Candidate Shortcut]
\label{def:lcdst_candidate_shortcut}
For a region $H$ and a vertex $u \in V(Q) \setminus V(A_H)$ on one of its separator paths $Q$, we define the \textbf{candidate shortcut} of $u$ to be the path $P_u = (a_u,\ldots,u)$ from some $a_u$ minimizing
\begin{align*}
P_u & = \argmin_{P = (a,\ldots,u)}{\left [ \frac{d_{T}(r,a)}{\epsilon h} + w_H(P) \right ]} = \argmin_{P = (a,\ldots,u)}{\left [ \frac{d_{T}(r,a)}{\epsilon h} + \frac{l(P)}{\epsilon h} + \frac{c(P)}{\lpvalue\numlevels} \right ]},
\end{align*}
where the minimum is taken over all ancestor vertices $a \in V(A_H)$ and paths $P$ from $a$ to $u$ in $H$.
\end{definition}
\noindent
Recall $d_{T}(r,a_u)$ is precisely the distance under $l$ from $r$ to the ancestor vertex $a_u$ in our current solution $T$.
After constructing the separator hierarchy, we buy the union of all the separator edges and set their costs to zero for the remainder of the algorithm. The cost function $c$ and our weight functions $w_H$ use these modified costs, whereas $\lpvalue$ remains its original value.



\paragraph{Step 3: Purchasing Shortcuts.}
We now define how we decide, for each vertex $u$ on the separator, whether to buy its candidate shortcut $P_u$. First, we immediately choose not to buy $P_u$ if it is ``far away'' from the root; namely, we skip $u$ if the minimum it achieves in \Cref{def:lcdst_candidate_shortcut} is more than $1 + 1/\epsilon$, as described below.
\begin{definition}[Skipped Vertex]
\label{def:lcdst_skipped_vertex}
For a region $H$ and a vertex $u \in V(Q) \setminus V(A_H)$ on one of its separator paths $Q$, we call $u$ a \textbf{skipped vertex} if its candidate shortcut $P_u$ from some ancestor vertex $a_u$ to $u$ satisfies
\begin{align*}
\frac{d_{T}(r,a_u)}{\epsilon h} + w_H(P_u) & > 1 + \frac{1}{\epsilon},
\end{align*}
or if no such candidate shortcut exists.
\end{definition}
\noindent
 In the above definition, $T$ is the solution at the point when we consider vertex $u$. If $u$ is a skipped vertex, then we don't buy its candidate shortcut and move on to the next vertex on the separator. Informally, for a non-skipped vertex $u$, we buy its candidate shortcut $P_u$ when the length improvement $P_u$ provides to reach $u$ as compared to reusing previous shortcuts is at least the cost of $P_u$ (scaled appropriately). We formalize this as follows.
\begin{definition}[Purchased Shortcut]
\label{def:lcdst_purchased_shortcut}
For a region $H$ and a vertex $u \in V(Q) \setminus V(A_H)$ on one of its separator paths $Q$, we purchase its candidate shortcut $P_u$ from $a_u$ to $u$ when
\begin{align*}
\frac{d_{T}(r,u) - [d_{T}(r,a_u) + l(P_u)]}{\epsilon h} & \geq \frac{c(P_u)}{\lpvalue\numlevels}.
\end{align*}
\end{definition}
\noindent
In this case, we call $P_u$ a \textbf{purchased shortcut} and add its edges to our solution $T$. Here, $d_{T}(r,u)$ is the distance from $r$ to $u$ in $T$ \emph{before} buying $P_u$, i.e.\ using a previously purchased shortcut to an earlier separator vertex $v$ and traveling along the separator $Q$ from $v$ to $u$. If $u$ is the first non-skipped vertex on the separator $Q$, we always buy $P_u$, so its candidate shortcut is always a purchased shortcut. \Cref{alg:lcdst_buyshortcuts} describes the pseudocode for buying these shortcuts in each region of the separator hierarchy.

\begin{algorithm}[H]
    \caption{BuyShortcuts($T,H, A_H$)}    
    \begin{algorithmic}[1]
        \label{alg:lcdst_buyshortcuts}
        \State If $\lpvalue = 0$, set $T \gets T \cup E_H$ \Return
        \For{$Q \in \{Q_1, Q_2, Q_3\}$}
            \For{$u \in V(Q) \setminus V(A_H)$ in order along the directed separator path $Q$}
                \State Compute $P_u$ from $a_u$ to $u$ as the candidate shortcut using \Cref{def:lcdst_candidate_shortcut}
                \If{$u$ is a skipped vertex using \Cref{def:lcdst_skipped_vertex}}
                    \State \textbf{continue}
                \EndIf
                \If{$P_u$ is a purchased shortcut using \Cref{def:lcdst_purchased_shortcut}}
                    \State $T \gets T \cup P_u$
                \EndIf
            \EndFor
        \EndFor
    \end{algorithmic}
\end{algorithm}
\noindent Finally, we compute a shortest path tree of $T$ rooted at $r$ under $l$. The pseudocode of our LC-DST rounding algorithm is described in \Cref{alg:lcdst}, which uses \Cref{alg:lcdst_buyseparators,alg:lcdst_buyshortcuts} as subroutines. It remains to show that our algorithm's solution $T$ reaches every terminal $t \in S$ within distance at most $(1+\epsilon)h$ from $r$ and that its cost is at most $O\left(\frac{\log^2k}{\epsilon}\right)$ times that of the fractional solution $(x,f)$.

\begin{algorithm}[H]
    \caption{LC-DST Rounding Algorithm}    
    \begin{algorithmic}[1]
        \label{alg:lcdst}
        \setlength{\itemsep}{1pt}
        \Statex \textbf{Input}: $G = (V,E)$, root $r$, terminals $S$, and fractional solution $(x,f)$.
        \State Remove edges with $x_e = 0$; $T \gets \emptyset$
        \State \textsc{BuySeparators}$(T, G,\{r\})$; set separator costs to zero 
        \For{region $H$ in order of the separator hierarchy starting at the root region}
            \State \textsc{BuyShortcuts}$(T, H, A_H)$ 
        \EndFor
        \State \Return shortest path tree of $T$ rooted at $r$ under $l$
    \end{algorithmic}
\end{algorithm}



\subsection{Charging Cost to Length}
In this section, we prove the following bound on the total cost of the solution returned by our algorithm.
\begin{restatable}[Total Cost]{lemma}{lemTotalCost}\label{lem:lcdst_total_cost}
\Cref{alg:lcdst} constructs a subgraph $T$ whose cost satisfies
\begin{align*}
c(T) & \leq O\left( \frac{\log^{2}{k}}{\epsilon} \right) \sum_{e \in E} c(e)x_e.
\end{align*}
\end{restatable}
\noindent
To do so, first we will bound the costs of the separator edges we buy in each region of the separator hierarchy (see \Cref{def:lcdst_separator_hierarchy}). Then, we bound the costs of the shortcuts we buy for each separator in a region by charging their costs to the successive improvements in length they provide to reach the final vertex of the separator from the root. Finally, we sum the costs of our solution in each region across the separator hierarchy.

\begin{figure}[t]
    \centering
    \includegraphics[width=0.5\linewidth]{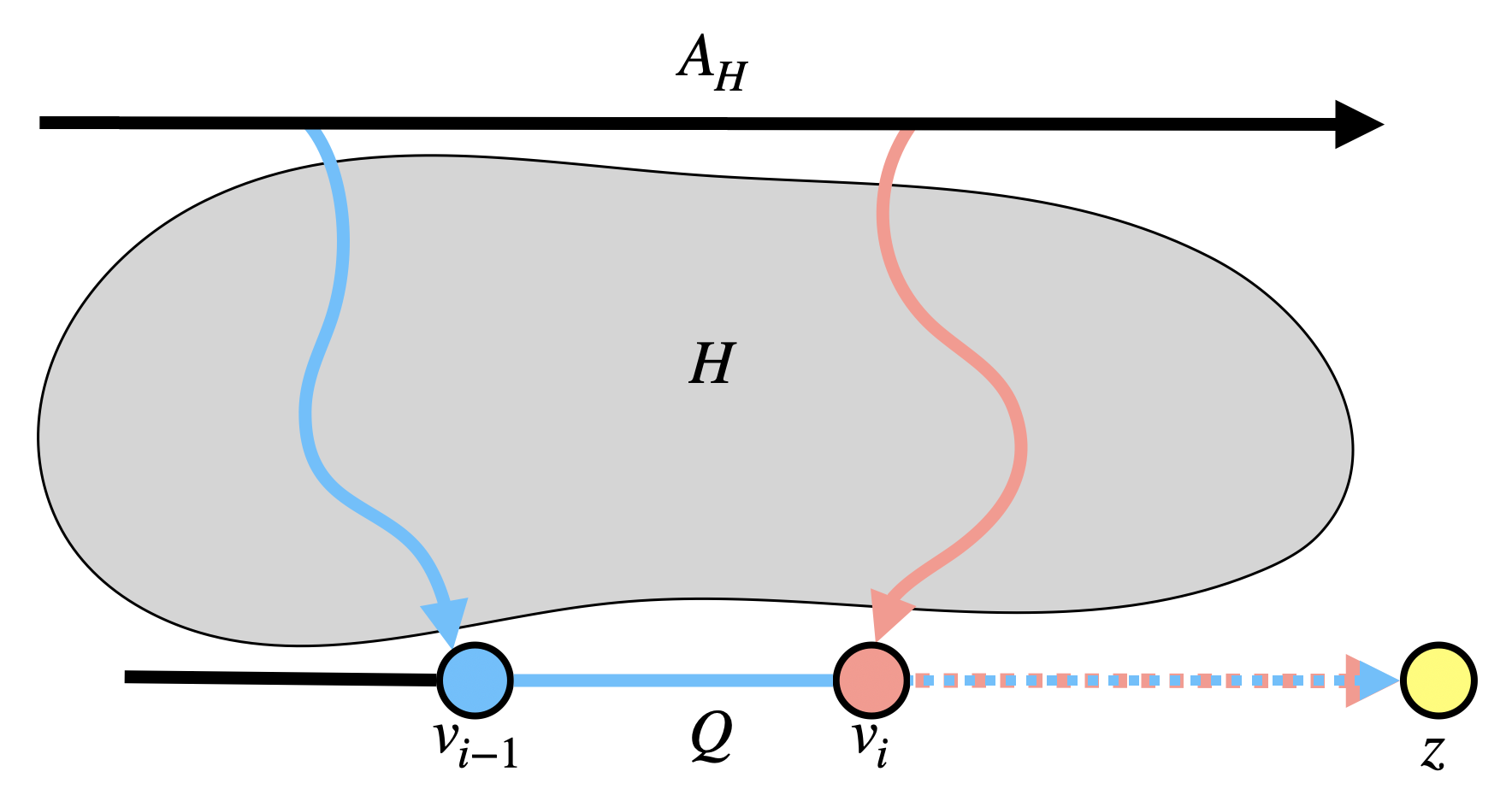}
    \caption{Charging shortcut cost to improvements in length. In the region $H$, there are two paths from $A_H$ to $z$: one through $v_{i-1}$ (blue) and the other through $v_{i}$ (red). They share the suffix $Q[v_i,z]$ of the separator.}
    \label{fig:costlength}
\end{figure}

We bound the cost and length of each separator path as follows.
\begin{lemma}[Separator Cost and Length]
\label{lem:lcdst_separator_cost}
Let $H$ be a region of the separator hierarchy with $\lpvalue > 0$, and let $Q$ be a separator path chosen in $H$ by \Cref{alg:lcdst_buyseparators}. Then,
\begin{align*}
c(Q) \leq O\left(\frac{\log{k}}{\epsilon}\right)\lpvalue,\qquad l(Q) & \leq (1+\epsilon)h.
\end{align*}
\end{lemma}
\begin{proof}
We achieve a bound on both the cost and the length by first observing that $Q$ is a shortest path under the hierarchical mixture metric $w_H$ and then applying the pruning radius of \Cref{def:lcdst_pruning} to $Q$. Since $Q$ is a shortest path under $w_H$ and we retained its vertices after pruning with radius $1 + 1/\epsilon$, we have
\begin{align*}
\frac{l(Q)}{\epsilon h} + \frac{c(Q)}{\lpvalue\numlevels} & = w_{H}(Q) \leq 1 + \frac{1}{\epsilon}.
\end{align*}
Note that the lengths $l$ and the costs $c$ are nonnegative, so dropping the length term gives $c(Q) \leq O\left(\frac{\log{k}}{\epsilon}\right)\lpvalue$ and dropping the cost term gives $l(Q) \leq (1+\epsilon)h$, as desired. 
\end{proof}
\noindent
We now use this $O(h)$ length bound of the separator to pay for the cost of the shortcuts. In the below bound on the cost of the shortcuts, $c$ assigns each separator edge to have cost 0; we have already bounded their cost above.
\begin{lemma}[Shortcut Cost]
\label{lem:lcdst_shortcut_cost}
Let $H$ be a region of the separator hierarchy with $\lpvalue > 0$, and let $P_{1},\ldots,P_{m}$ be the purchased shortcuts for a separator path $Q$ in $H$. Then,
\begin{align*}
\sum_{i=1}^{m} c(P_i) & \leq O\left(\frac{\log{k}}{\epsilon} \right)\lpvalue.
\end{align*}
\end{lemma}
\begin{proof}
Let $Q = (q_0, \ldots, z)$. We charge the cost of each shortcut after the first to the improvement in length it provides to reach $z$ (the final vertex of $Q$).

We index the purchased shortcuts in the order our algorithm purchases them as $P_{i} = (a_i, \ldots, v_i)$. To measure the length of the path to reach $z$ from the root $r$ through the shortcut $P_i$, define
\begin{align*}
L_i & := d_{T}(r,a_i) + l(P_i) + l(Q[v_i,z]),
\end{align*}
where $T$ is our solution immediately before buying shortcut $P_i$. We fix $L_i$ thereafter. For $i \geq 2$, write $\Delta_i = L_{i-1} - L_{i}$ for the decrease in the length to reach $z$ from $r$ after purchasing $P_i$.

Next, we show that the length improvement $\Delta_i$ pays for the cost of $P_i$. Since $P_{i}$ is a purchased shortcut, we have by the definition of a purchased shortcut (\Cref{def:lcdst_purchased_shortcut}) that
\begin{align*}
\frac{c(P_i)}{\lpvalue\numlevels} & \leq \frac{d_{T}(r,v_i) - [d_T(r,a_i) + l(P_i)]}{\epsilon h},
\end{align*}
where again $T$ is our solution before buying $P_i$. Observe that an existing path to $v_i$ from $r$ before we bought $P_{i}$ is the one that follows a path from $r$ to $a_{i-1}$, takes the previous shortcut $P_{i-1}$, and travels along the subpath of the separator $Q[v_{i-1},v_{i}]$. The $r$--$z$ path using $P_{i-1}$ and the $r$--$z$ path using $P_{i}$ both share the suffix $Q[v_{i},z]$ (see \Cref{fig:costlength}). Adding the length of this shared suffix to the above inequality gives
\begin{align*}
\frac{c(P_i)}{\lpvalue\numlevels}  \leq \frac{d_{T}(r,v_i) + l(Q[v_i,z]) - L_i}{\epsilon h} 
 \leq \frac{L_{i-1} - L_{i}}{\epsilon h} 
 = \frac{\Delta_i}{\epsilon h}.
\end{align*}
This also shows $\Delta_i \geq 0$. Now that each new shortcut pays for itself through the improvement in length it provides to reach $z$, we bound their total cost by summing these improvements.

We sum the charges for all shortcuts after the first, and with intermediate path lengths canceling, we get
\begin{align*}
\sum_{i=2}^{m} \frac{c(P_i)}{\lpvalue\numlevels} & \leq \frac{\sum_{i=2}^{m} \Delta_{i}}{\epsilon h} = \frac{L_1 - L_m}{\epsilon h}.
\end{align*}
Since the lengths are nonnegative, $L_{m} \geq 0$. Thus, to bound the sum of the costs, it suffices to bound $L_1$. Since $v_{1}$ was not skipped, we have by the definition of a skipped vertex (\Cref{def:lcdst_skipped_vertex}) that
\begin{align*}
\frac{d_{T}(r,a_1) + l(P_1)}{\epsilon h} + \frac{c(P_1)}{\lpvalue\numlevels} & \leq 1 + \frac{1}{\epsilon},
\end{align*}
where $T$ is the solution just before purchasing $P_1$. Thus, we have that 
\begin{align}
d_{T}(r,a_1) + l(P_1) & \leq (1+\epsilon)h \leq O(h), \label{eq:sc_cost_1}
\end{align}
and 
\begin{align}
\frac{c(P_1)}{\lpvalue\numlevels} & \leq O(1/\epsilon). \label{eq:sc_cost_3}
\end{align}
Note that after buying $P_1$, we can reach $z$ from $r$ by following the path from $r$ to $a_1$, taking the shortcut $P_1$ to $v_1$, and traveling along the suffix of the separator $Q[v_1,z]$. Using \Cref{lem:lcdst_separator_cost}, we have that 
\begin{align}
l(Q[v_1,z]) & \leq O(h). \label{eq:sc_cost_2}
\end{align}
Adding the length bounds in \eqref{eq:sc_cost_1} and \eqref{eq:sc_cost_2} together, we have that $L_1 \leq O(h)$. Combining this with the cost bound in \eqref{eq:sc_cost_3}, we have
\begin{align*}
\frac{\sum_{i=1}^{m} c(P_i)}{\lpvalue\numlevels} & \leq O(1/\epsilon) + \frac{L_1 - L_m}{\epsilon h} \leq O(1/\epsilon).
\end{align*}
This gives
\begin{align*}
\sum_{i=1}^{m} c(P_i) & \leq O\left(\frac{\numlevels}{\epsilon}\right)\lpvalue,
\end{align*}
as desired.
\end{proof}

\begin{lemma}[Total Cost]
\label{lem:lcdst_total_cost} 
    \Cref{alg:lcdst} constructs a subgraph $T$ whose cost satisfies
\begin{align*}
c(T) & \leq O\left( \frac{\log^{2}{k}}{\epsilon} \right) \sum_{e \in E} c(e)x_e.
\end{align*}
\end{lemma}
\begin{proof}
For any region $H$, \Cref{lem:lcdst_separator_cost,lem:lcdst_shortcut_cost} give a combined cost of $O(\log{k}/\epsilon)\lpvalue$ per separator and the shortcuts we buy for it in $H$. We have at most three separators per region, so our total cost per region is still $O(\log{k}/\epsilon)\lpvalue$. Observe that the edge sets of the regions in any one of our recursive levels are disjoint by definition of separators and the separator hierarchy. Therefore, the sum of $\lpvalue$ across the regions of any one recursive level is at most $\sum_{e \in E} c(e)x_e$. Summing over $\numlevels$ levels gives
\begin{align*}
c(T) & \leq O\left(\frac{\log{k}}{\epsilon} \right ) \sum_{H} \lpvalue  \leq O\left(\frac{\log^{2}{k}}{\epsilon} \right ) \sum_{e \in E} c(e)x_e. \qedhere
\end{align*}
\end{proof}

\subsection{Charging Length to Cost}
In this section, we prove the following lemma that bounds the distance from the root to each terminal in our solution.

\begin{figure}[t]
    \centering
    \includegraphics[width=0.3\linewidth]{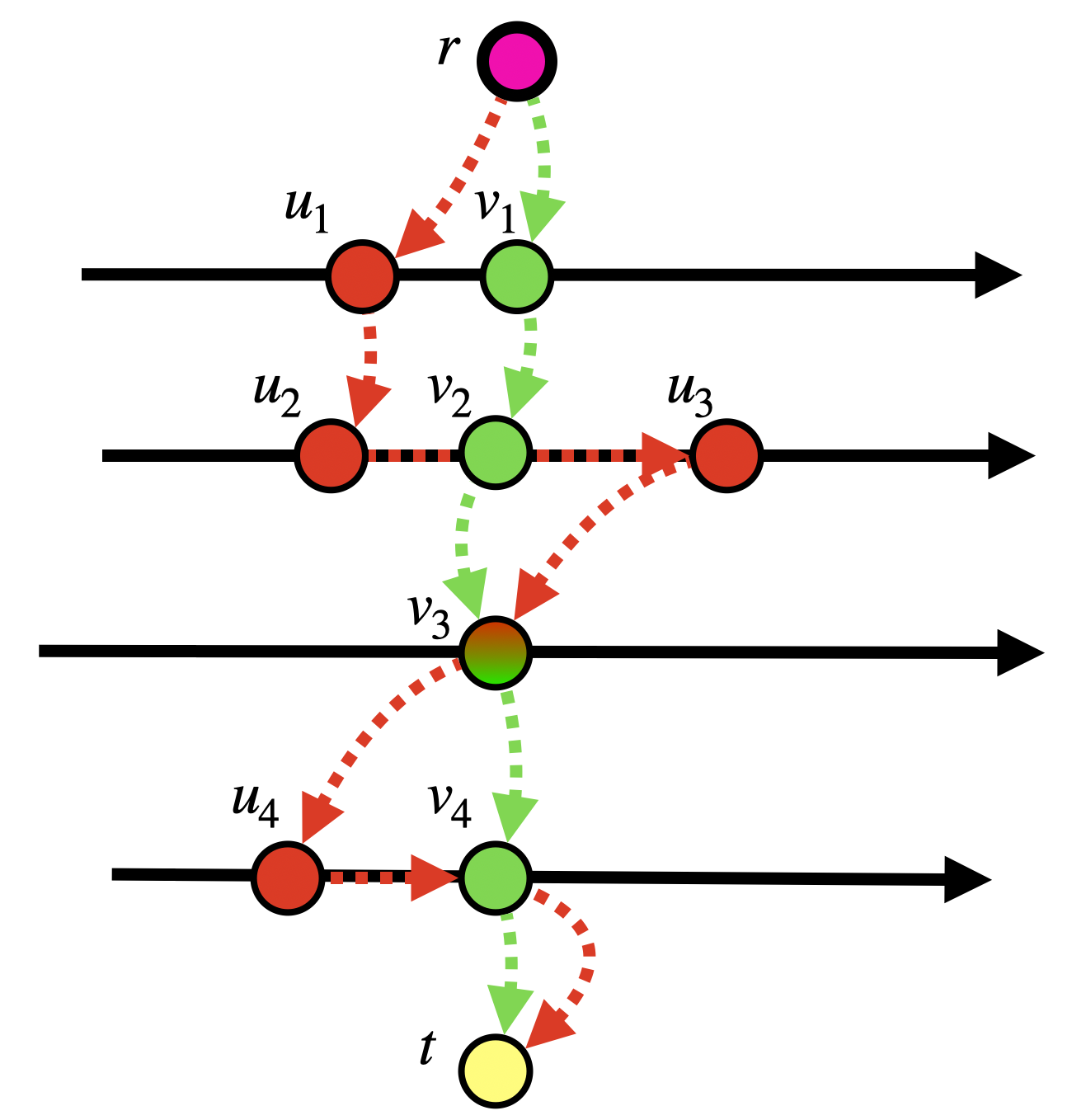}
    \caption{Charging length to canonical path cost. Pictured is a canonical $r$-$t$ path containing vertices $(r,v_1,v_2,v_3,v_4,t)$ in the dotted green edges crossing a set of separator paths in the horizontal solid black edges. Our algorithm's $r$-$t$ path containing vertices $(r,u_1,u_2,v_2,u_3,v_3,v_4,v_4)$ is given in the dotted red edges, and can deviate from the canonical path based on where each purchased shortcut originates and ends for every parent/child separator pair. 
    }
    \label{fig:lengthcost}
\end{figure}

\begin{restatable}[Total Length]{lemma}{lemTotalLength}\label{lem:lcdst_total_length}
\Cref{alg:lcdst} constructs a subgraph $T$ that, for every terminal $t \in S$, satisfies 
\begin{align*}
\dalg(r,t) \leq (1+\epsilon)h.
\end{align*}
\end{restatable}
\noindent For each terminal $t \in S$, we first find a path from $r$ to $t$ of length at most $h$ and which, on average, has cost proportional to the LP cost on each region of the separator hierarchy. We then charge the increase in length of our solution as compared to this path to the latter's cost. We define such a path formally as follows.
\begin{definition}
\label{def:lcdst_canonical_def}
A \textbf{canonical path} for a terminal $t$ is a directed path $P^{*}_{t} = (r,\ldots, t)$ in $G$ satisfying
\begin{align*}
l(P^{*}_{t}) & \leq h, \qquad \sum_{H : t \in I_H} \frac{c(P^{*}_{t} \cap E_H)}{\lpvalue\numlevels} \leq 1,
\end{align*}
where the sum is over regions of the separator hierarchy with $t$ as an interior terminal and $P^{*}_{t} \cap E_H$ are the edges of $P^{*}_{t}$ in $E_{H}$ (see \Cref{def:lcdst_separator_hierarchy}).
\end{definition}
\noindent 
We now prove that such a canonical path exists.
\begin{lemma}[Canonical Path Exists]
\label{lem:lcdst_canonical_exists}
Every terminal $t \in S$ has a canonical path as defined in \Cref{def:lcdst_canonical_def}.
\end{lemma}
\begin{proof}
We choose a canonical path minimizing the sum of the above cost terms in each region among the paths carrying positive flow in the LP solution $(x,f)$. Let $f^{(t)}_{P}$ denote the flow assigned to $P$ for terminal $t$ in $(x,f)$. Choose
\begin{align*}
P^{*}_{t} & = \argmin_{\substack{P \in \mathcal{P}^{(h)}(r,t) \\ f^{(t)}_{P} > 0}} \sum_{H : t \in I_H} \frac{c(P \cap E_H)}{\lpvalue\numlevels}.
\end{align*}
The LP routes at least one unit of flow to $t$, so such a path exists. Every path in $\mathcal{P}^{(h)}(r,t)$ has length at most $h$, so $l(P^{*}_{t}) \leq h$.

Now, to prove the cost bound, we will show that on average, a path carrying positive flow achieves the desired bound and use this to bound the cost terms of $P^{*}_{t}$. In each region $H$ of the separator hierarchy, the cost of $t$'s flow is at most the cost of the LP restricted to that region. Namely,
\begin{align*}
\sum_{P \in \mathcal{P}^{(h)}(r,t)} f^{(t)}_{P} c(P \cap E_H) & = \sum_{e \in E_H} c(e) \sum_{\substack{P \in \mathcal{P}^{(h)}(r,t) \\ e \in P}} f^{(t)}_{P} \leq \sum_{e \in E_H} c(e)x_e = \lpvalue.
\end{align*}
We now combine these cost bounds in each region to bound the total cost of the paths carrying flow to $t$. We sum over the regions with $t \in I_H$. There are at most $\numlevels$ such regions, so
\begin{align}
\sum_{H : t \in I_H} \sum_{P \in \mathcal{P}^{(h)}(r,t)} f^{(t)}_{P} \frac{c(P \cap E_H)}{\lpvalue\numlevels} & \leq 1, \label{eq:lcdst_averaging}
\end{align}
where regions with $\lpvalue = 0$ contribute $0$ to this sum. 

It remains to transfer this bound from the LP flow to the chosen path $P^{*}_{t}$. Interchanging the sums in \eqref{eq:lcdst_averaging} groups together all the cost terms of each region for each path with positive flow. Since $P^{*}_{t}$ minimizes the sum of the cost terms in each region among paths with positive flow, we obtain
\begin{align*}
\left (\sum_{P \in \mathcal{P}^{(h)}(r,t)} f^{(t)}_{P} \right )\sum_{H : t \in I_H} \frac{c(P^{*}_{t} \cap E_H)}{\lpvalue\numlevels} & \leq \sum_{P \in \mathcal{P}^{(h)}(r,t)} f^{(t)}_{P} \left (\sum_{H : t \in I_H} \frac{c(P \cap E_H)}{\lpvalue\numlevels} \right ) \leq 1.
\end{align*}
The LP sends at least one unit of flow to $t$, so
\begin{align*}
\sum_{H : t \in I_H} \frac{c(P^{*}_{t} \cap E_H)}{\lpvalue\numlevels}  \leq \dfrac{1}{\left (\sum\limits_{P \in \mathcal{P}^{(h)}(r,t)} f^{(t)}_{P} \right )}  \leq 1.
\end{align*}
Together with $l(P^{*}_{t}) \leq h$ from before, this proves that there exists a canonical path for $t$.
\end{proof}
\noindent
We have shown that each terminal has a canonical path of length at most $h$ with a small sum of cost terms for each region of the separator hierarchy (see \Cref{def:lcdst_separator_hierarchy}). To use this path to bound the length of our solution, we will compare our shortcuts with subpaths of the canonical path in each region. However, to make this comparison in each region, we must first show that pruning retains these subpaths throughout the separator hierarchy as follows.
\begin{definition}[Canonical Segment]
\label{def:lcdst_canonical_segment}
For any $t \in S$, let $P^{*}_{t}$ be its canonical path. For a region $H$ with $t \in I_H$, let $a_H$ be the last vertex of $P^{*}_{t}$ in the ancestor subgraph $A_H$; let $v_H$ be the last vertex of $P^{*}_{t}$ in $A_H \cup \mathcal{Q}$. Then, the \textbf{canonical segment} of $P^{*}_{t}$ in $H$ is 
\begin{align*}
P^{*}_{t,H} & := P^{*}_{t}[a_H,v_H].
\end{align*}
\end{definition}
\noindent
In the above, the last vertex is the one that appears last in the order $P^{*}_{t} = (r,\ldots, t)$; $\mathcal{Q}$ is the separator chosen in $H$, and $P^{*}_{t}[a_H,v_H]$ is the directed subpath from $a_H$ to $v_H$.
\begin{lemma}[Canonical Path Survives Pruning]
\label{lem:lcdst_canonical_survives}
For any $t \in S$, let $P^{*}_{t}$ be its canonical path. Each canonical segment $P^{*}_{t,H}$ is retained in $\operatorname{prune}(H)$ as defined in \Cref{def:lcdst_pruning}.
\end{lemma}
\begin{proof}
We proceed by induction from parent regions to child regions (that is, the height of the hierarchy) to show that the vertices of $P^{*}[a_H,t]$ lie in $H$\emph{ before }pruning. This holds at the root, where $a_H = r$ and this suffix is the entire canonical path $P^{*}_{t}$.

Now, for the inductive step, consider any region $H$ satisfying our inductive hypothesis. Let $u$ be any vertex on the suffix $P^{*}_{t}[a_H, t]$. We use the definition of $P^{*}_{t}$ as the canonical path to bound the distance under $w_H$ of $u$ from the contracted root as described in \Cref{def:lcdst_pruning}. Using the definition of $P^{*}_{t}$ as the canonical path in \Cref{def:lcdst_canonical_def}, we have
\begin{align*}
w_{H}(P^{*}_{t}[a_H,u]) & \leq \frac{l(P^{*}_{t}[a_H,u])}{\epsilon h} + \frac{c(P^{*}_{t}[a_H,u])}{\lpvalue\numlevels} \\
& \leq \frac{l(P^{*}_{t})}{\epsilon h} + \frac{c(P^{*}_{t} \cap E_H)}{\lpvalue\numlevels} \\
& \leq \frac{1}{\epsilon} + 1,
\end{align*}
where the second inequality follows from the inductive hypothesis and the last inequality follows from the length and cost bounds of a canonical path, by definition. 

Also, observe that this bounds the distance of $u$ from the contracted root. After contracting $A_H$ in $H \cup A_H$, the prefix $P^{*}_{t}[a_H,u]$ gives a path from the contracted root to $u$ of mixture metric distance at most $1 + 1/\epsilon$.

It remains to show that the inductive hypothesis holds for the next region, i.e.\ that the remaining suffix $P^{*}_{t}[v_H,t]$ lies in the child region of $H$ containing $t$. If $v_H = t$, then the canonical path ends on the separator and we are done. Otherwise, all vertices after $v_H$ lie outside of $A_H \cup \mathcal{Q}$, by definition of $v_H$ in \Cref{def:lcdst_canonical_segment}. They therefore lie in the same connected component as $t$, which contains $P^{*}_{t}[v_H,t]$, and we are done.
\end{proof}
\noindent
We have shown that the canonical path for any terminal survives pruning. Thus, we can now compare our candidate shortcuts (see \Cref{def:lcdst_candidate_shortcut}) with the canonical segments in each region of the separator hierarchy. This lets us obtain a path to each segment's endpoint that is no longer than that of the canonical path, apart from an increase paid for by that segment's cost.
\begin{lemma}[Charging Excess Length to Canonical Path Cost]
\label{lem:lcdst_charge_canonical}
For any $t \in S$, let $P^{*}_{t,H} = [a,v]$ be one of its canonical segments. After \Cref{alg:lcdst_buyshortcuts} decides whether to buy the candidate shortcut of $v$, our solution contains an $r$--$v$ path $P$ satisfying
\begin{align*}
l(P) & \leq d_{T}(r,a) + l(P^{*}_{t,H}) + \epsilon h \frac{c(P^{*}_{t,H})}{\lpvalue\numlevels},
\end{align*}
where we assume $v$ is not skipped by \Cref{alg:lcdst_buyshortcuts} (see \Cref{def:lcdst_skipped_vertex}) and $T$ is our solution immediately before making a decision on $v$'s candidate shortcut.
\end{lemma}
\begin{proof}
We first compare our algorithm's candidate shortcut with that of the canonical segment in order to bound the length of the path from $r$ to $v$ that we obtain.

Let $P_{v} = (a_{v},\ldots,v)$ be the candidate shortcut for $v$ (\Cref{def:lcdst_candidate_shortcut}). The canonical segment $P^{*}_{t}$ lies in $H$ after pruning and starts at $a \in V(A_H)$ by \Cref{lem:lcdst_canonical_survives}. Because $P_v$ is a shortest path under $w_H$, accounting for the distance from $r$ to reach each starting ancestor, we have
\begin{align*}
\frac{d_{T}(r,a_v)}{\epsilon h} + w_{H}(P_v) & \leq \frac{d_{T}(r,a)}{\epsilon h} + w_{H}(P^{*}_{t,H}) \leq \frac{d_{T}(r,a)}{\epsilon h} + \frac{l(P^{*}_{t,H})}{\epsilon h} + \frac{c(P^{*}_{t,H})}{\lpvalue\numlevels}.
\end{align*}
Multiplying the above by $\epsilon h$ gives
\begin{align}
d_{T}(r,a_v) + \epsilon h w_{H}(P_v) & \leq d_{T}(r,a) + l(P^{*}_{t,H}) + \epsilon h\frac{c(P^{*}_{t,H})}{\lpvalue\numlevels}. \label{eq:spineq}
\end{align}
Recall that in \Cref{alg:lcdst_buyshortcuts}, the costs $c$ are modified with separator edges having cost 0. But, replacing these costs with the original costs can only result in a larger upper bound on the right-hand side; so, the inequality in \eqref{eq:spineq} holds for original costs on the right-hand side as well.

We now show that regardless of whether $P_v$ was a purchased shortcut (\Cref{def:lcdst_purchased_shortcut}), this gives an $r$--$v$ path of length at most $d_{T}(r,a_v) + \epsilon h w_{H}(P_v)$ by charging to the canonical path; see \Cref{fig:lengthcost}.

\smallskip
\noindent \textbf{Case 1: We buy $P_v$.}
We follow a shortest path in $T$ from $r$ to $a_v$ under $l$, then take the shortcut $P_v$ to $v$. This gives a path $P$ in the solution after purchasing $P_v$ satisfying
\begin{align*}
l(P) & \leq d_T(r,a_v)+l(P_v) \leq d_T(r,a_v)+\epsilon h\,w_H(P_v) \leq d_{T}(r,a) + l(P^{*}_{t,H}) + \epsilon h\frac{c(P^{*}_{t,H})}{\lpvalue\numlevels}.
\end{align*}

\smallskip
\noindent \textbf{Case 2: We do not buy $P_v$.}
In this case, if we do not purchase the shortcut $P_v$, consider the shortest path $P$ from $r$ to $v$ in $T$ that already existed. If we did not purchase $P_v$, it must have been because $P$ already reaches $v$ with length comparable to the length of the canonical segment, up to a factor proportional to the cost of the latter. We formalize this as follows.

Since we assume $v$ is not a skipped vertex, it must be that $P_v$ was not a purchased shortcut according to \Cref{def:lcdst_purchased_shortcut}. This happens precisely when
\begin{align*}
\frac{d_T(r,v) - (d_T(r,a_v) + l(P_v))}{\epsilon h} & < \frac{c(P_v)}{\lpvalue\numlevels},
\end{align*}
where notably $T$ is our solution before buying $P_v$, i.e.\ it reaches $v$ by using a previous shortcut and traveling along the separator. With some rearranging, we arrive at
\begin{align*}
d_T(r,v) & < d_T(r,a_v)+\epsilon h\,w_H(P_v) \leq d_{T}(r,a) + l(P^{*}_{t,H}) + \epsilon h\frac{c(P^{*}_{t,H})}{\lpvalue\numlevels}.
\end{align*}
Thus, the additional length of our algorithm's path as compared to the canonical path can be charged to the cost of the latter.
In either case, we have
\begin{align*}
l(P) & \leq d_{T}(r,a) + l(P^{*}_{t,H}) + \epsilon h\frac{c(P^{*}_{t,H})}{\lpvalue\numlevels}.\qedhere
\end{align*}
\end{proof}\noindent

\noindent
For any terminal $t$, we can now apply \Cref{lem:lcdst_charge_canonical} in each region of the separator hierarchy to bound our total root-to-terminal distances as follows.

\lemTotalLength*
\begin{proof}
Fix $t \in S$. We show by induction from parent regions to child regions that the additional length of our algorithm as compared to the canonical path for $t$ can be charged proportionally to the cost of the canonical path.

Using \Cref{lem:lcdst_canonical_exists}, we can find a canonical path $P^{*}_{t}$ for $t$; number its canonical segments in parent-to-child order as $P^{*}_{t,H_i} = P^{*}_{t}[v_{i-1}, v_{i}]$ in region $H_{i}$ for $1 \leq i \leq m$. Namely, we show that after \Cref{alg:lcdst_buyshortcuts} returns when called with region $H_i$, our solution satisfies
\begin{align*}
d_{T}(r,v_i) & \leq l(P^{*}_{t}[r,v_i]) + \epsilon h \sum_{j=1}^{i} \frac{c(P^{*}_{t,H_j})}{X_{H_{j}}\numlevels} \leq (1+\epsilon)h,
\end{align*}
where $T$ is our solution at that time. Since \Cref{lem:lcdst_charge_canonical} assumes that the endpoints of the canonical segments $P^{*}_{t,H}$ of $P^{*}_{t}$ are not skipped as defined by \Cref{def:lcdst_skipped_vertex}, we also verify this in our induction.

For $i = 0$, both sides are 0. Otherwise, suppose $i \geq 1$. Immediately before we decide whether to purchase the candidate shortcut for $v_i$ using \Cref{def:lcdst_purchased_shortcut}, the inductive hypothesis bounds the distance to the ancestor vertex $v_{i-1}$. Reaching $v_{i-1}$ from $r$ and following $P^{*}_{t,H_i}$ from $v_{i-1}$ to $v_{i}$ gives
\begin{align*}
d_{T}(r,v_{i-1}) + l(P^{*}_{t,H_i}) + \epsilon h \frac{c(P^{*}_{t,H_i})}{X_{H_i}\numlevels} & \leq l(P^{*}_{t}[r,v_{i}]) + \epsilon h\sum_{j=1}^{i} \frac{c(P^{*}_{t,H_j})}{X_{H_j}\numlevels} \\
& \leq l(P^{*}_{t}) + \epsilon h\sum_{H : t\in I_H} \frac{c(P^{*} \cap E_H)}{X_H\numlevels},
\end{align*}
where the second line uses $P^{*}_{t,H_j} \subseteq P^{*}_{t} \cap E_{H_{j}}$. Now, since $P^{*}_{t}$ is a canonical path, the definition of a canonical path by \Cref{def:lcdst_canonical_def} gives
\begin{align*}
l(P^{*}_{t}) + \epsilon h\sum_{H : t\in I_H} \frac{c(P^{*} \cap E_H)}{X_H\numlevels} & \leq h + \epsilon h \cdot 1 \leq (1 + \epsilon)h.
\end{align*}
Since the canonical segments $P^{*}_{t,H_i}$ is a possible candidate shortcut (\Cref{def:lcdst_candidate_shortcut}) for $v_i$, this bound also implies that the vertex $v_i$ is not skipped by \Cref{alg:lcdst_buyshortcuts}.

We can now apply \Cref{lem:lcdst_charge_canonical} to charge the additional length of our solution as compared to the canonical path to the cost of the canonical path. After deciding whether to buy the candidate shortcut for $v_i$, the solution contains an $r$--$v_i$ path $P$ satisfying
\begin{align*}
l(P) & \leq d_T(r,v_{i-1}) + l(P^{*}_{t,H_i}) + \epsilon h\frac{c(P^*_{t,H_i})}{X_{H_i}\numlevels} \leq l(P_t^*[r,v_i]) + \epsilon h \sum_{j=1}^{i} \frac{c(P^*_{t,H_j})}{X_{H_j}\numlevels}.
\end{align*}
Here, $T$ is the solution immediately before deciding whether to buy the candidate shortcut for $v_i$; the first inequality comes from \Cref{lem:lcdst_charge_canonical} and the second inequality comes from the inductive hypothesis. Then, since $P$ is in $T$ after the decision, we have that $d_{T}(r,v_i) \leq l(P)$, completing the inductive step.

Now, we can consider the final vertex of $P^{*}_{t} = (r,\ldots,t)$ with $v_m = t$ to bound the root-to-terminal length of our algorithm overall. The above inequality then gives
\begin{align*}
d_{T}(r,t) & \leq l(P^{*}_{t}) + \epsilon h \sum_{j=1}^{m} \frac{c(P^{*}_{t,H_j})}{X_{H_{j}}\numlevels} \leq (1+\epsilon)h,
\end{align*}
where the second inequality once again comes from the definition of $P^{*}_{t}$ as a canonical path.
\end{proof}

\subsection{Proof of \Cref{thm:LCDST}}
\thmLCDST*
\begin{proof}
\Cref{lem:lcdst_total_cost,lem:lcdst_total_length} give us the desired cost and length bounds, respectively. We take a shortest path tree under $l$, which cannot increase cost and preserves the root-to-terminal distances. There are at most $\log{k}$ levels of the separator hierarchy since each child has at most half its parent's terminals, there are at most $O(k\log{k})$ regions of the separator hierarchy, each of which requires polynomial time. The restricted paths subroutine in the separation oracle for \ref{lp:lcdst} takes time $\poly(n,1/\epsilon)$. Thus, our LC-DST rounding algorithm runs in time $\poly(n,1/\epsilon)$.
\end{proof}


\section{LP-Competitive Length-Constrained Steiner Forest}\label{sec:lcsf}

\newcommand{\candidatealg}{\textsf{ConstructCandidates}}
\newcommand{\lcsfalg}{\textsf{LCSF-Algo}}

In this section, we prove the following lemma reducing, up to polylogarithmic factors, length-constrained Steiner Forest to 
length-constrained Steiner Tree in planar graphs.

\begin{lemma}
\label{lem:lcst_to_lcsf_main}
  Suppose, for any $\epsilon > 0$, there exists an algorithm $\mathcal A$ that is an LP-competitive $\alpha$-approximation with $1 + \epsilon/4$ length slack for 
  LC-ST in planar graphs. Then, there exists an LP-competitive $O\left(\frac{\log^3 k}{\epsilon}(\alpha + \log k)\right)$-approximation with $1 + \epsilon$ length slack 
  for LC-SF in planar graphs that runs in time $\poly(n,1/\epsilon), T(\mathcal A(n, \epsilon))$, where $T(\mathcal A(n, \epsilon))$ is the runtime of $\mathcal A$.
\end{lemma}
\noindent This, combined with \Cref{thm:LCDST}, concludes the proof of \Cref{thm:LCSF}, restated below. 

\thmLCSF*

\subsection{Algorithm}

We describe the algorithm below; the formal description is given by \Cref{alg:lcsf}.
Let $(x, f)$ be an optimal solution to \ref{lp:lcsf}. We let $X := \sum_{e \in E} c(e) x_e$ denote the cost of this fractional solution.
Let $\mathcal P$ denote the set of paths in $\cup_{i \in [k]} \mathcal P^{(h)}(s_i, t_i)$ with non-zero $f_p$, and let $\mathcal P_i^h := \mathcal P^{(h)}(s_i, t_i) \cap \mathcal P$. By the discussion in \Cref{sec:prelim}, $(x, f)$ has polynomial support, thus $\mathcal P$ has polynomially many paths.
Throughout the algorithm, $D'$ denotes the set of remaining uncovered pairs, $F$ denotes the current partial solution, and $\mathcal H$ denotes the set of candidates. It is not difficult to verify that the algorithm runs in time $\poly(n, 1/\epsilon)$.

\begin{algorithm}[h]
\caption{LC-SF Rounding Algorithm}
\label{alg:lcsf}
\begin{algorithmic}
  \setlength{\itemsep}{1pt}
  \Statex \textbf{Input:} $G = (V, E)$, demand pairs $D$, and fractional solution $(x, f)$.
  \Statex
  \State $X:= \sum_{e \in E} c(e) x_e$
  \State $F := \emptyset, D' := D$, $\mathcal H = \emptyset$
  \For{$i \in [k]$}
    \If{$\exists p \in \mathcal P^h_i$ with $c(p) \leq \frac X k$}
      \State $F \gets F \cup p$
      \State $D' \gets D' \setminus \{(s_i, t_i)\}$
    \EndIf
  \EndFor
  \State Remove all paths $p \in \cup_{i \in [k]} \mathcal P^h_i$ with $c(p) \geq 4X$, delete corresponding flow variables 
  \State $(x, f) \gets \frac 4 3 (x, f)$
  \Statex
  \State $\tau := \frac 1 \epsilon \log k$
  \For{$j = 1, \dots, \lceil \log(4k)/\log\tau \rceil$}
    \State $B_j = \left\{p \in \mathcal P: c(p) \in \left(\frac X k \tau^{j-1}, \frac X k \tau^j\right]\right\}$
    \State $\mathcal H \gets \mathcal H \cup \candidatealg(G, D', B_j, (x,f), j)$
  \EndFor
  \While{$D' \neq \emptyset$}
    \State Let $(H, D_{H}) \in \mathcal H$ be a candidate minimizing $c(H)/|D' \cap D_{H}|$
    \State $D' \gets D' \setminus D_H$
    \State $F \gets F \cup E(H)$
  \EndWhile
  \State \Return $F$
\end{algorithmic}
\end{algorithm}
\noindent
Recall the high-level approach discussed in \Cref{sec:tech_overview}. We construct \emph{candidate junction trees} via length-constrained separators. Our goal is to divide each separator path $Q$ into chunks of length $O(\epsilon h)$, which behave as roots in our LC-ST instances. To do so, we want to argue that for each $(s_i, t_i)$ pair, all flow paths of $\mathcal P_i^h$ intersect $Q$ within a small subpath; thus, $s_i$-$t_i$ flow is spread across only a small number of chunks. This requires choosing an appropriate cost parameter on which to define the mixture metric for the separator. Suppose the mixture metric is defined with cost parameter $\gamma$ and length $h$. If a flow path $p$ has cost much larger than $\gamma$, then its intersections with $Q$ can span a long subpath and thus intersect too many chunks. On the flip side, if the cost of flow paths intersecting a chunk is much smaller than $\gamma$, then we cannot adequately charge the cost of buying the chunk to the cost of the paths intersecting it. Therefore, we start by grouping flow paths into buckets of similar cost, and choosing the mixture metric separately for each bucket.

\paragraph{Step 1: Preprocessing and Bucketing Paths.}
While the \emph{average} cost of paths in each $\mathcal P_i^h$ is at most $X$, individual paths that carry small amounts of flow may have much larger cost. Therefore, path costs can range from $0$ to $|E|\max_{e \in E} c(e)$. Naively bucketing would lead to $O(\log (n\max_{e} c(e)))$ buckets. Therefore, we first preprocess paths in $\mathcal P$ to restrict costs to $(X/k, 4X)$, reducing the number of buckets.

First, for each $i \in [k]$, if there exists a path of cost at most $\frac X k$, we buy the path and remove the demand pair. 
Next, we throw away all paths of cost at least $4X$, and scale the remaining fractional solution by $\frac 4 3$. The following lemma demonstrates that high-cost paths constitute a small amount of the total flow between demand pairs. In particular, the scaled fractional solution supports a flow of $1$ between demand pairs.

\begin{lemma}
\label{lem:lcsf_preprocessing}
  For $(s_i, t_i) \in D$, $\sum_{p \in \mathcal P^h_i: c(p) < 4X} f_p \geq \frac 3 4$.
\end{lemma}
\begin{proof}
  Fix $(s_i, t_i) \in D$. The average weighted cost of paths in $\mathcal P_i^h$ is at most 
  \[\sum_{p \in \mathcal P_i^h} f_p \sum_{e \in p} c(e) = \sum_{e \in E} c(e) \sum_{p \in \mathcal P_i^h: e \in p} f_p \leq \sum_{e \in E} c(e) x_e = X.\]
Therefore, $4X \sum_{p \in \mathcal P_i^h: c(p) \geq 4X} f_p \leq X$, so $\sum_{p \in \mathcal P_i^h: c(p) \geq 4X} f_p \leq \frac 1 4$. 
\end{proof}
\noindent
Following the preprocessing, all remaining paths in $\mathcal P$ have cost strictly between $\frac X k$ and $4X$. We bucket paths based on their cost. Let $B_j$ denote the set of all paths in $\mathcal P$ with $\frac X k \tau^{j-1} < c(p) \leq \frac X k \tau^j$, where $\tau = \frac {\log k} \epsilon$. For each bucket $B_j$, let $\gamma_j := \frac X k \tau^j$, and let $w_j$ denote the mixture metric with respect to cost parameter $\gamma_j$ and length parameter $h$; i.e. 
\[w_j(e) = c(e) + \frac X {hk} \tau^j \cdot l(e).\]
The following lemma is a direct consequence of \Cref{lem:mixture}, since each path $p \in B_j$ has $l(p) \leq h$ and $c(p) \leq \gamma_j$. 

\begin{lemma}
\label{lem:lcsf_bucket_mixture}
  For every path $p \in B_j$, the mixture cost $w_j(p)$ is at most $2\gamma_j$.
\end{lemma}

\paragraph{Step 2: Constructing Candidate Rooted Solutions.} For each bucket $B_j$, we construct a set $\mathcal H_j$ of \emph{candidate} rooted partial solutions. Fix a bucket $B_j$. We follow a divide-and-conquer algorithm using the length-constrained separator given by \Cref{lem:lc-separator}. From an arbitrary root $r \in V$, we find three shortest paths (with respect to the mixture metric $w_j$) $Q_1, Q_2, Q_3$ from $r$ in $G$. For each $Q \in \{Q_1, Q_2, Q_3\}$, we do the following. 

Subdivide $Q$ into \emph{chunks}: these are subpaths of $Q$ each of mixture cost $\frac \epsilon 4 \gamma_j$. We write $\mathcal C_Q$ to denote the set of all chunks of $Q$. Following \Cref{lem:mixture}, observe that for each chunk $C \in \mathcal C_Q$, $c(C) \leq \frac \epsilon 4 \gamma_j$, and $l(C) \leq \frac \epsilon 4 h$.
Next, we assign each path $p \in B_j$ to every chunk $C$ that it intersects. We let $\mathcal P_C$ denote the paths assigned to $C$, and let $G_C$ denote the subgraph of $G$ consisting of $C$ and all paths assigned to $C$.

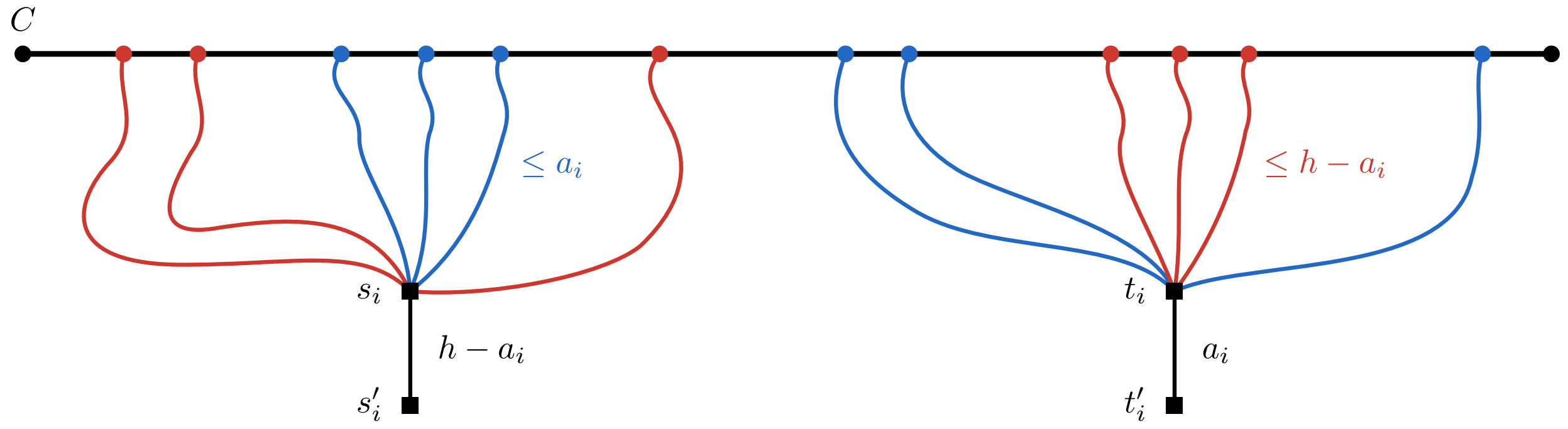
\begin{figure}[t]
    \centering
\begingroup%
\pgfmathsetlengthmacro{\rootedxunit}{(\linewidth-12pt)/14.4}%
\definecolor{flowblue}{RGB}{35,105,195}%
\definecolor{flowred}{RGB}{205,55,45}%
\begin{tikzpicture}[
  x=\rootedxunit,y=1cm,
  line cap=round,line join=round,
  flow/.style={line width=1.2pt},
  vertex/.style={inner sep=0pt,outer sep=0pt,minimum size=5pt},
  every node/.style={font=\normalfont\normalsize,inner sep=1pt}
]
  \draw[line width=1.7pt] (0,0) -- (14.4,0);
  \node[vertex,circle,fill=black] at (0,0) {};
  \node[vertex,circle,fill=black] at (14.4,0) {};
  \node[anchor=base] at ([yshift=7pt]0,0) {$C$};

  \coordinate (s) at (3.65,-2.50);
  \coordinate (t) at (10.85,-2.50);
  \coordinate (sdummy) at (3.65,-3.70);
  \coordinate (tdummy) at (10.85,-3.70);

  \draw[flow,flowred] (s)
    .. controls (3.20,-2.00) and (2.55,-2.23) .. (1.45,-2.22)
    .. controls (.34,-2.20) and (.48,-1.55) .. (.82,-1.14)
    .. controls (1.14,-.75) and (.86,-.40) .. (.95,0);
  \draw[flow,flowred] (s)
    .. controls (3.30,-1.70) and (2.62,-1.70) .. (1.85,-1.83)
    .. controls (1.12,-1.98) and (1.42,-1.35) .. (1.59,-1.03)
    .. controls (1.83,-.66) and (1.54,-.34) .. (1.65,0);
  \draw[flow,flowblue] (s)
    .. controls (3.60,-1.77) and (3.16,-1.24) .. (3.17,-.88)
    .. controls (3.18,-.47) and (2.78,-.38) .. (3.00,0);
  \draw[flow,flowblue] (s)
    .. controls (3.90,-1.82) and (3.74,-1.23) .. (3.83,-.84)
    .. controls (3.96,-.47) and (3.61,-.34) .. (3.80,0);
  \draw[flow,flowblue] (s)
    .. controls (4.25,-1.97) and (4.42,-1.27) .. (4.53,-.84)
    .. controls (4.65,-.43) and (4.38,-.34) .. (4.50,0);
  \draw[flow,flowred] (s)
    .. controls (4.38,-2.57) and (5.47,-2.34) .. (5.82,-2.02)
    .. controls (6.26,-1.55) and (6.28,-1.12) .. (6.08,-.72)
    .. controls (5.94,-.42) and (5.82,-.25) .. (6.00,0);
  \node[text=flowblue,anchor=base west] at (4.65,-1.25) {$\leq a_i$};

  \draw[flow,flowblue] (t)
    .. controls (10.35,-1.90) and (9.10,-2.13) .. (8.40,-1.65)
    .. controls (7.65,-1.15) and (7.55,-.60) .. (7.75,0);
  \draw[flow,flowblue] (t)
    .. controls (10.57,-1.84) and (9.40,-1.57) .. (8.85,-1.25)
    .. controls (8.30,-.90) and (8.20,-.40) .. (8.35,0);
  \draw[flow,flowred] (t)
    .. controls (10.62,-1.81) and (10.25,-1.24) .. (10.35,-.86)
    .. controls (10.45,-.48) and (10.12,-.31) .. (10.25,0);
  \draw[flow,flowred] (t)
    .. controls (10.94,-1.82) and (10.83,-1.26) .. (10.96,-.84)
    .. controls (11.10,-.46) and (10.75,-.31) .. (10.90,0);
  \draw[flow,flowred] (t)
    .. controls (11.27,-1.87) and (11.45,-1.22) .. (11.52,-.81)
    .. controls (11.64,-.44) and (11.39,-.28) .. (11.55,0);
  \draw[flow,flowblue] (t)
    .. controls (11.50,-2.15) and (13.45,-2.38) .. (13.65,-1.30)
    .. controls (13.80,-.75) and (13.65,-.35) .. (13.75,0);
  \node[text=flowred,anchor=base west] at (11.65,-1.25) {$\leq h-a_i$};

  \foreach \x in {.95,1.65,6.00,10.25,10.90,11.55}
    \node[vertex,circle,fill=flowred] at (\x,0) {};
  \foreach \x in {3.00,3.80,4.50,7.75,8.35,13.75}
    \node[vertex,circle,fill=flowblue] at (\x,0) {};

  \draw[line width=1.2pt] (s) -- (sdummy);
  \draw[line width=1.2pt] (t) -- (tdummy);
  \node[anchor=base west] at ([xshift=7pt,yshift=-3pt]3.65,-3.10)
    {$h-a_i$};
  \node[anchor=base west] at ([xshift=7pt,yshift=-3pt]10.85,-3.10)
    {$a_i$};

  \foreach \p in {s,t,sdummy,tdummy}
    \node[vertex,rectangle,fill=black] at (\p) {};

  \foreach \p/\lab in {s/s_i,t/t_i,sdummy/s_i',tdummy/t_i'}
    \node[anchor=east,text height=1.5ex,text depth=.5ex]
      at ([xshift=-7pt]\p) {$\lab$};
\end{tikzpicture}%
\endgroup%
    \caption{Constructing an LC-ST instance for chunk $C$ and terminal pair $(s_i, t_i)$, with dummy terminals $(s_i', t_i')$ and corresponding edges. At least half the $s_i$-$C$ flow is on blue paths of length at most $a_i$, and at least half the $C$-$t_i$ flow is on red paths of length at most $h-a_i$.}
    \label{fig:lcsf_rooted_instance}
\end{figure}

Our goal is to reduce to length-constrained Steiner Tree by using the chunks as ``roots''. Naively, one could assign a pair $(s_i, t_i)$ to chunk $C$ if sufficient $s_i$-$t_i$ flow travels through paths in $\mathcal P_C$. Running the given length-constrained Steiner Tree algorithm on this instance would result in paths from $C$ to $s_i, t_i$, each of length at most $(1+\epsilon') h$ (we will choose $\epsilon' = \epsilon/4$). We make two changes to this simple approach. 
\begin{itemize}
  \item The distance from $s_i$ to $t_i$ would be as large as $(2 + \epsilon') h$ from the LC-ST algorithm, in addition to the length of the chunk. To achieve our desired $(1 + \epsilon)$ length slack, we instead find the \emph{median} distance $a_i$ that $s_i$ takes to reach chunk $C$ along paths in $\mathcal P_i^h \cap \mathcal P_C$, weighted by the corresponding $f_p$ values. That is, $a_i$ satisfies 
  \[\min\left(\sum_{p \in \mathcal P_i^h \cap \mathcal P_C: l(p[s_i,C]) \leq a_i} f_p, \sum_{p \in \mathcal P_i^h \cap \mathcal P_C: l(p[C,t_i]) \leq h-a_i} f_p \right) \geq \frac 1 2 \sum_{p \in \mathcal P_i^h \cap \mathcal P_C} f_p.\]
  We aim to use $a_i$ as the length constraint for $s_i$, and $h - a_i$ as the length constraint for $t_i$. Since the given algorithm for LC-ST does not allow for non-uniform length constraints, we add a dummy node $s_i'$, connected to $s_i$ via an edge of length $h-a_i$, and a dummy node $t_i'$, connected to $t_i$ via an edge of length $a_i$. Setting $s_i', t_i'$ as the terminals in the LC-ST instance results in our desired length constraints. See \Cref{fig:lcsf_rooted_instance}.
  \item With this approach, guaranteeing that each $(s_i, t_i)$ pair is covered by at least one chunk would require scaling $x_e$ up by a factor of $O(\log^2 k /\log\tau)$. This is because $s_i$-$t_i$ flow could be spread evenly among the $O(\log k/\log \tau)$ buckets $B_j$, and also evenly among the $O(\log k)$ recursion levels. This can be wasteful. For example, suppose every pair sends $\frac 1 {q}$ flow through $q$ chunks, where $q$ is the total number of chunks. If we were to buy candidate solutions corresponding to all chunks, the flow in each chunk would need to be scaled up by $q$, which could result in a $q^2$ multiplicative loss in cost. However, it suffices to buy only the cheapest candidate. Similarly, if each pair sends constant flow through $O(1)$ chunks, we could scale by a small constant factor.

  Intuitively, our approach is to consider all possible scaling factors for all possible candidate solutions, and find the best one. Formally, we construct candidate solutions for different \emph{flow thresholds}. For each threshold $\beta = 1, \frac 1 2, \frac 1 4, \dots, \frac{\epsilon}{M\log^2k}$, we consider all pairs that send at least $\beta$ flow through $C$, and construct a candidate rooted solution. Here, $M$ is a large constant specified in \Cref{lem:lcsf_flow_coverage}. We will show in \Cref{sec:lcsf_analysis} that this approach allows us to offset the scaling cost via a weighted averaging argument.
\end{itemize}

\noindent
This concludes the construction of candidate solutions rooted at chunks in $Q$. We recurse on connected components of $G \setminus (Q_1 \cup Q_2 \cup Q_3)$ until all pairs are separated; this occurs in $O(\log k)$ levels of recursion. See \Cref{alg:lcsf_candidates} for a formal description of the candidate construction;
$\mathcal A$ denotes the given approximation algorithm for LC-ST in planar graphs.

\begin{algorithm}[h]
\caption{$\candidatealg(G, D, B_j, (x,f), j)$}
\label{alg:lcsf_candidates}
\begin{algorithmic}
  \setlength{\itemsep}{1pt}
  \If{$|D| = 0$} \Return $\emptyset$ \EndIf
  \State $\mathcal H_j \gets \emptyset$
  \State $Q_1 \cup Q_2 \cup Q_3 \gets$  separator given by \Cref{lem:lc-separator} on $G$ with arbitrary root $r \in V$ and edge-weights $w_j$
  \For{$Q \in \{Q_1, Q_2, Q_3\}$}
    \State Subdivide $Q$ into chunks of mixture cost $\frac \epsilon 4 \gamma_j$ each; let $\mathcal C_Q$ denote the set of chunks
    \For{each chunk $C \in \mathcal C_Q$}
      \State $\mathcal P_{C} \gets \{p \in B_j: p \cap C \neq \emptyset\}$
      \State $G_C \gets C \cup \left(\cup_{p \in \mathcal P_{C}} p\right)$
      \For{each $(s_i, t_i) \in D$ with $\sum_{p \in \mathcal P_i^h \cap \mathcal P_C} f_p > 0$}
        \State $a_i \gets$ median distance from $s_i$ to $C$ along paths of $P_i^h \cap \mathcal P_C$, weighted by $f_p$
        \State Add dummy vertices $s_i', t_i'$ and $0$-cost edges $(s_i, s_i'), (t_i, t_i')$ to $G_C$
        \State Set edge lengths $l(s_i, s_i') = h-a_i$ and $l(t_i, t_i') = a_i$
      \EndFor
      \For{$\beta = 1, \frac 1 2, \frac 1 4, \dots , \frac \epsilon {M \log^2 k}$}
        \State $D_{C, \beta} \gets \{(s_i, t_i) \in D: \sum_{p \in \mathcal P_i^h \cap \mathcal P_C} f_p \geq \beta \}$
        \If{$D_{C, \beta} \neq \emptyset$}
            \State Obtain $G_C'$ by contracting $C$ to a root $r_C$ in $G_C$
            \State $T_{C, \beta} \gets \mathcal A(G_C', r_C, \{s_i', t_i' : (s_i, t_i) \in D_{C, \beta}\})$
            \State Obtain $H_{C, \beta}$ by uncontracting $C$ in $T_{C, \beta}$ and deleting dummy vertices/edges
            \State $\mathcal H_j \gets \mathcal H_j \cup \{(H_{C, \beta}, D_{C, \beta})\}$
        \EndIf
      \EndFor
    \EndFor
  \EndFor
  \For{connected component $S$ of $G \setminus (Q_1 \cup Q_2 \cup Q_3)$}
    \State $B_{j,S} \gets \{p \in B_j: p \subseteq S\}$
    \State $D_S \gets \{(s_i, t_i) \in D: s_i, t_i \in V(S)$
    \State $\mathcal H_j \gets \mathcal H_j \cup \candidatealg(S, D_S, B_{j,S}, (x, f), j)$
  \EndFor
  \Return $\mathcal H_j$
\end{algorithmic}
\end{algorithm}

\paragraph{Step 3: Choosing Min-Density Solutions.}
The final step of \Cref{alg:lcsf} is to greedily buy the candidate with lowest \emph{density}; this is the candidate minimizing the ratio of cost over new demand pairs covered. We repeat this process until all terminal pairs are covered.

\subsection{Cost and Length Analysis}
\label{sec:lcsf_analysis}
\paragraph{Length Bound.}
We argue that each candidate solution satisfies the length constraints. 
\begin{lemma}
  \label{lem:lcsf_length}
  For each candidate solution $(H, D_H) \in \mathcal H$, for all $(s_i, t_i) \in D_H$, there exists an $s_i$-$t_i$ path in $H$ of length at most $(1 + \epsilon) h$.
\end{lemma}
\begin{proof}
  Fix a candidate solution $(H, D_H) \in \mathcal H$, and let $C, \beta$ be the corresponding chunk and flow threshold. Recall that $H_{C, \beta}$ is formed by uncontracting $C$ from $T_{C, \beta}$, which is given by the LC-ST algorithm $\mathcal A$. By the guarantees of $\mathcal A$, $T_{C, \beta}$ contains a path from the dummy vertex $s_i'$ to the contracted root $r_C$ of length at most $(1 + \epsilon') h$. Since the only path to $s_i'$ in $T_{C, \beta}$ is via $s_i$, this implies an $r_C$-$s_i$ path of length at most $(1 + \epsilon')h - l(s_i, s_i') = a_i + \epsilon' h$. Similarly, $T_{C, \beta}$ contains an $r_C$-$t_i$ path of length at most $(1 + \epsilon')h - l(t_i, t_i') = (h - a_i) + \epsilon' h$. Uncontracting $C$ yields an $s_i$-$t_i$ path in $H$ of length at most 
  \[(a_i + \epsilon' h) + (h-a_i + \epsilon' h) + l(C) \leq h + (2\epsilon' + \epsilon/4)h.\]
  Choosing $\epsilon' = \epsilon/4$ concludes the proof.
\end{proof}

\paragraph{Cost Analysis.}
We bound the cost of the solution given by \Cref{alg:lcsf} by proving the following lemma. 
\begin{lemma}
    \label{lem:lcsf_cost_main}
    Let $F$ be the solution given by \Cref{alg:lcsf}. Then, the cost of $F$ is at most $O\left(\frac{\log^3 k}\epsilon(\alpha + \log k)\right) X$, where $\alpha$ is the approximation ratio of the LC-ST algorithm $\mathcal A$.
\end{lemma}
\noindent 
First, observe that the total cost of the paths bought in the preprocessing step is at most $k(X/k) = X$. To bound the cost of chosen candidates, we will argue that at each iteration, there exists a low density choice. In particular, we will show that for every remaining set $D'$ of demand pairs, there exists a candidate rooted solution satisfying 
\[\frac {c(H)}{|D_H \cap D'|} \leq \frac{O(\frac 1 \epsilon \log^2 k(\alpha + \log k))X}{|D'|}.\] 
By standard set-cover arguments, this implies that the total cost of chosen candidate solutions is at most 
$O\left(\frac{\log^3 k}\epsilon (\alpha + \log k)\right)X$, as desired.

For the remainder of this section, we demonstrate that there always exists a candidate solution of low density. 
We do so via a weighted averaging argument. The following lemma is simple.
\begin{lemma}
    Let $a_1, \dots, a_k \in \mathbb R_{\geq 0}$ and $b_1, \dots, b_k \in \mathbb R_{\geq 0}$, and choose any weights $\lambda_1, \dots, \lambda_k \in \mathbb R_{\geq 0}$. Then 
    \[\min_{i \in [k]} \frac {a_i}{b_i} \leq \frac{\sum_{i \in [k]} \lambda_i a_i} {\sum_{i \in [k]} \lambda_i b_i}.\]
\end{lemma}
\noindent
We give each candidate $(H_{C, \beta}, D_{C, \beta})$ a weight of $4\beta$. By the preceding lemma, it suffices to show the following.
\begin{enumerate}
    \item For every remaining set $D'$ of demand pairs, $\sum_{C, \beta} 4\beta |D_{C, \beta} \cap D'| \geq |D'|$, and 
    \item The total weighted cost is bounded by $\sum_{C, \beta} 4\beta c(H_{C, \beta}) \leq O\left(\frac{\log^2 k}\epsilon(\alpha + \log k)\right)X$.
\end{enumerate}
\noindent
The first point follows from \Cref{lem:lcsf_flow_coverage}, which states that for any terminal pair, the sum of weights of candidates covering that pair is at least $1$. The proof of the second point is given by \Cref{lem:lcsf_total_cost}.
We start with a technical lemma charging each vertex of $G$ to a small number of chunks.
It will be helpful in to consider the set of \emph{all} chunks, across separators, levels of recursion, and buckets; we denote this set as $\mathcal C$. 

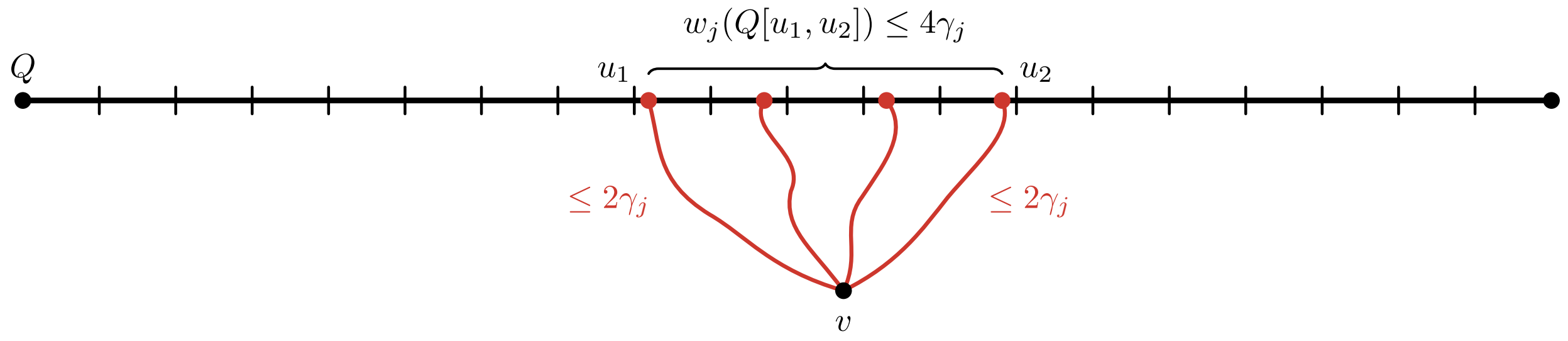
\begin{figure}[t]
    \centering
\begingroup%
\pgfmathsetlengthmacro{\chargingxunit}{(\linewidth-12pt)/16}%
\definecolor{chargingflow}{RGB}{205,55,45}%
\begin{tikzpicture}[
  x=\chargingxunit,y=1cm,
  line cap=round,line join=round,
  flow/.style={draw=chargingflow,line width=1.2pt},
  vertex/.style={circle,inner sep=0pt,outer sep=0pt,minimum size=5pt},
  every node/.style={font=\normalfont\normalsize,inner sep=1pt}
]
  \draw[line width=1.7pt] (0,0) -- (16,0);
  \foreach \k in {1,...,19} {
    \pgfmathsetmacro{\x}{.8*\k}
    \draw[line width=.85pt]
      ([yshift=-4pt]\x,0) -- ([yshift=4pt]\x,0);
  }
  \node[vertex,fill=black] at (0,0) {};
  \node[vertex,fill=black] at (16,0) {};
  \node[anchor=base] at ([yshift=7pt]0,0) {$Q$};

  \begin{scope}[shift={(2.4,0)}]
  \coordinate (u1) at (4.15,0);
  \coordinate (qleft) at (5.36,0);
  \coordinate (qright) at (6.64,0);
  \coordinate (u2) at (7.85,0);
  \coordinate (v) at (6.19,-2.00);

  \draw[flow] (v)
    .. controls (5.43,-1.78) and (5.17,-1.42) .. (4.79,-1.20)
    .. controls (4.21,-.84) and (4.28,-.47) .. (u1);
  \draw[flow] (v)
    .. controls (5.87,-1.56) and (5.55,-1.35) .. (5.64,-.95)
    .. controls (5.81,-.57) and (5.17,-.33) .. (qleft);
  \draw[flow] (v)
    .. controls (6.38,-1.60) and (6.15,-1.33) .. (6.38,-1.02)
    .. controls (6.61,-.67) and (6.89,-.33) .. (qright);
  \draw[flow] (v)
    .. controls (6.77,-1.71) and (7.02,-1.35) .. (7.28,-1.02)
    .. controls (7.59,-.65) and (7.99,-.31) .. (u2);

  \node[text=chargingflow,anchor=base east] at (4.21,-1.13)
    {$\leq 2\gamma_j$};
  \node[text=chargingflow,anchor=base west] at (7.66,-1.13)
    {$\leq 2\gamma_j$};

  \foreach \p in {u1,qleft,qright,u2}
    \node[vertex,fill=chargingflow] at (\p) {};
  \node[vertex,fill=black] at (v) {};
  \node[anchor=base] at ([yshift=-12pt]v) {$v$};

  \node[anchor=base east] at ([xshift=-4pt,yshift=7pt]u1) {$u_1$};
  \node[anchor=base west] at ([xshift=4pt,yshift=7pt]u2) {$u_2$};
  \draw[decorate,decoration={brace,amplitude=3pt},line width=.75pt]
    ([yshift=8pt]u1) -- ([yshift=8pt]u2)
    node[midway,above=8pt] {$w_j(Q[u_1,u_2])\leq 4\gamma_j$};
  \end{scope}
\end{tikzpicture}%
\endgroup%
    \caption{Red paths denote all $p \in \cup_{C \in \mathcal C_Q} \mathcal P_C$ containing $v$. The mixture cost of each of these paths is at most $2\gamma_j$, giving us a bound on the mixture cost of $Q[u_1, u_2]$.}
    \label{fig:lcsf_charging}
\end{figure}

\begin{lemma}
    \label{lem:lcsf_charging}
    For each $v \in V$, $v \in V(\mathcal P_C)$ for at most 
    $O(\log^2 k/(\epsilon \log \tau))$ chunks in $\mathcal C$. In particular, each edge $e \in E$ is in $G_C$ for at most $O(\log^2 k/(\epsilon \log \tau))$ chunks.
\end{lemma}
\begin{proof}
    Fix a vertex $v \in V$, a bucket $B_j$ and a separator path $Q$. Let $\mathcal C_Q$ denote the set of all chunks of $Q$. Consider the set of all points $u \in V(Q)$ such that there exists a path $p \in \cup_{C \in \mathcal C_Q} \mathcal P_C$ with $u,v \in V(p)$. Let $u_1, u_2 \in V(Q)$ be the closest and furthest such points from $r$. This implies that $v \in V(\mathcal P_C)$ only for chunks with non-empty intersection with $Q[u_1, u_2]$. 

    Recall, by \Cref{lem:lcsf_bucket_mixture}, that all paths in $B_j$ (and thus all paths in $\mathcal P_C$ for all $C \in \mathcal C_Q$) have mixture cost at most $2\gamma_j$. Let $d_j$ be the distance function corresponding to the shortest-path metric given by $w_j$. Since there exists a path $p \in \mathcal P_C$ with $v, u_1 \in V(p)$, $d_j(v, u_1) \leq 2\gamma_j$. Similarly, $d_j(v,u_2) \leq 2\gamma_j$. Since $d_j$ is a metric, $d_j(u_1, u_2) \leq 4\gamma_j$. In particular, $w_j(Q[u_1, u_2]) \leq 4\gamma_j$, since $Q$ is a shortest path with respect to $w_j$. Since each chunk has mixture cost $\frac \epsilon 4 \gamma_j$, there are at most $(4\gamma_j) (4/(\epsilon \gamma_j)) + 2 \leq (16/\epsilon) + 2$ chunks that intersect $Q[u_1, u_2]$. Therefore, $v \in V(\mathcal P_C)$ for at most $O(1/\epsilon)$ chunks of $\mathcal C_Q$. See \Cref{fig:lcsf_charging}.

    Multiplying over $3$ separator paths, $O(\log k)$ levels of recursion, and $O(\log k/\log\tau)$ buckets yields the desired bound. Note that the components on which \Cref{alg:lcsf_candidates} recurses are disjoint, so each vertex belongs to at most one component at each recursion depth.

    For each $e \in E$, if $e \in G_C$ for a chunk $C \in \mathcal C$, then either $e \in E(\mathcal P_C)$ or $e \in E$. By the above, if $e \in E(\mathcal P_C)$, then its endpoints must be in $V(\mathcal P_C)$; this can occur for at most $O(\log^2 k/(\epsilon \log \tau))$ chunks. At a level of recursion, chunks are edge-disjoint. Therefore, over $O(\log k)$ levels of recursion and $O(\log k/\log \tau)$ buckets, $e \in C$ for at most $O(\log^2 k/\log \tau)$ chunks.
\end{proof}

\begin{lemma}
    \label{lem:lcsf_flow_coverage}
    For each $(s_i, t_i) \in D'$ after the preprocessing stage, $\sum_{(C, \beta): (s_i, t_i) \in D_{C, \beta}} 4\beta \geq 1.$
\end{lemma}
\begin{proof}
    Fix $(s_i, t_i) \in D'$. For ease of notation, let $f_i(C) := \sum_{p \in \mathcal P_i^h \cap \mathcal P_C} f_p$, this is the amount of flow that $(s_i, t_i)$ sends through chunk $C$. Since the recursive process continues until all terminal pairs are separated, every path $p \in \mathcal P_i^h$ must intersect the separator at some level of recursion. Therefore, $\sum_{C \in \mathcal C} f_i(C) = \sum_{p \in \mathcal P_i^h} f_p \geq 1$, where the last inequality follows from \Cref{lem:lcsf_preprocessing}.

     Let $\beta_{\min} = \frac \epsilon {M\log^2 k}$ be the minimum flow threshold. By \Cref{lem:lcsf_charging}, there are at most $O(\log^2k/(\epsilon \log \tau)) = O(\log^2k/\epsilon)$ chunks such that $f_i(C) > 0$. 
    Therefore, we can choose $M$ large enough so that $\sum_{C: f_i(C) < \beta_{\min}} f_i(C) \leq \frac 1 2$. This implies that at least half the $s_i$-$t_i$ flow belongs to chunks $C \in \mathcal C$ where $f_i(C)$ is above the minimum flow threshold, and is therefore included in a candidate solution. For each such chunk, there exists a candidate solution $H_{C, \beta} \in \mathcal H$ such that $\beta \geq \frac 1 2 f_i(C)$. Therefore,
    
    \[\sum_{(C, \beta): (s_i, t_i) \in D_{C, \beta}} 4\beta 
    \geq 4 \sum_{C: f_i(C) \geq \beta_{\min}} \frac 1 2 f_i(C)\geq 1. \qedhere\]
    
\end{proof}
\noindent
To bound the total weighted cost, we split the cost of a candidate $H_{C, \beta}$ into the ``rooted'' portion coming from the LC-ST algorithm (this corresponds to $T_{C, \beta}$ in \Cref{alg:lcsf_candidates}), and cost of the chunks. The rooted portion is simple to bound, since candidate weights are chosen specifically to offset the scaling needed to obtain a feasible LC-ST LP solution. For the chunk cost, since edges of a chunk are not paid for by the LP, we charge its cost to the flow paths intersecting it. The cost bucketing is necessary here to lower bound the path costs with respect to the chunk cost. 
We analyze the cost of the rooted portion in \Cref{lem:lcsf_rooted_cost} and the cost of the chunks in \Cref{lem:lcsf_chunk_cost}. 

\begin{lemma}
    \label{lem:lcsf_rooted_cost}
    For each candidate $(H_{C, \beta}, D_{C, \beta})$, following the notation in \Cref{alg:lcsf_candidates}, we let $T_{C, \beta}$ denote the rooted solution in which $C$ is contracted. 
    Then, $\sum_{C, \beta} 4\beta c(T_{C, \beta}) \leq O\left(\alpha \frac {\log^2 k}\epsilon\right) \sum_{e \in E} c(e)x_e$.
\end{lemma}
\begin{proof}
    Fix a candidate $H_{C, \beta}$, and recall that $T_{C, \beta}$ is the result of $\mathcal A(G_C', r_C, \{s_i', t_i' : (s_i, t_i) \in D_{C, \beta}\})$, where $G_C'$ is obtained from $G_C$ by contracting $C$ into a root $r_C$. 

    For each $(s_i, t_i) \in D_{C, \beta}$, $\sum_{p \in \mathcal P_i^h \cap \mathcal P_C} f_p \geq \beta$. Recall that $a_i$ was chosen to be the weighted median distance that $s_i$ takes to reach $C$ along paths in $\mathcal P_i^h \cap \mathcal P_C$. Therefore, 
    \[\sum_{p \in \mathcal P_i^h \cap \mathcal P_C: l(p[s_i,C]) \leq a_i} f_p \geq \frac 1 2 \sum_{p \in \mathcal P_i^h \cap \mathcal P_C} f_p \geq \frac \beta 2.\]
    If we were to extend paths to reach $s_i'$ via the dummy edge $(s_i, s_i')$, then since $l(s_i, s_i') = h - a_i$, $f$ would support a flow of at least $\beta/2$ from $s_i'$ to $r_C$ via $h$-length-bounded paths in $G_C'$. An analogous argument shows that when extending paths to $t_i'$ via the dummy edge, $f$ supports a flow of at least $\beta/2$ via $h$-length-bounded paths in $G_C'$. Therefore, $\frac 2\beta x$ restricted to $G_C'$ supports a feasible LP solution to the LC-ST instance (see \ref{lp:lcdst}).
    
    By the guarantees of the LC-ST algorithm $\mathcal A$, $c(T_{C, \beta}) \leq \alpha \frac2\beta \sum_{e \in G_C'} c(e)x_e \leq \alpha \frac 2\beta \sum_{e \in G_C} c(e)x_e$, where the last inequality uses the fact that the dummy edges have $0$ cost. The total weighted cost is therefore at most 
    \begin{align*}
        \sum_{C, \beta} 4\beta c(T_{C, \beta}) \leq 8\alpha \sum_{C, \beta} \sum_{e \in G_C} c(e) x_e \leq 8\alpha \sum_{\beta} \sum_{e \in E} c(e)x_e |\{C: e \in G_C\}| \leq O\left(\alpha \frac{\log^2 k}{\epsilon \log \tau}\right) \sum_{\beta} \sum_{e \in E} c(e)x_e,
    \end{align*}
    where the last inequality follows from \Cref{lem:lcsf_charging}. There are $O\left(\log \frac{\log^2k}{\epsilon}\right) = O(\log \tau)$ flow thresholds, yielding the desired total cost of $O\left(\alpha \frac{\log^2k}\epsilon\right) \sum_{e \in E} c(e) x_e$.
\end{proof}

\begin{lemma}
    \label{lem:lcsf_chunk_cost}
    For any fixed chunk $C \in \mathcal C$, $\sum_{\beta: H_{C, \beta} \in \mathcal H} 4\beta c(C) \leq O(\log k) \sum_{e \in G_C} c(e)x_e$.
\end{lemma}
\begin{proof}
    Consider a fixed chunk $C \in \mathcal C$, and let $j$ be the bucket $B_j$ associated with chunk $C$.
    Recall that $c(C) \leq \frac \epsilon 4 \gamma_j$. Therefore, 
    \[\sum_{\beta: H_{C, \beta} \in \mathcal H} 4\beta c(C) 
    \leq \epsilon \gamma_j \sum_{\beta: H_{C, \beta} \in \mathcal H} \beta.\]
    Observe that if $H_{C, \beta} \in \mathcal H$, then (1) there exists a pair $(s_i, t_i) \in D'$ with $\sum_{p \in \mathcal P_i^h \cap \mathcal P_C} f_p \geq \beta$, and (2) $H_{C, \beta'} \in \mathcal H$ for all flow thresholds $\beta' \leq \beta$. 
    Let $\beta^*$ be the largest flow threshold for which 
    $H_{C, \beta^*} \in \mathcal H$. Then,
    \[\sum_{\beta: H_{C, \beta} \in \mathcal H} \beta \leq 
    \sum_{\beta' \leq \beta^*} \beta' \leq 2\beta^* \leq 2\max_{i} \sum_{p \in \mathcal P_i^h \cap \mathcal P_C} f_p.\]
    Bucketing guarantees that for all $p \in \mathcal P_C$, $c(p) > \gamma_j/\tau$, and thus $\gamma_j f_p < \tau c(p)f_p$. Therefore,
    \begin{align*}
        \sum_{\beta: H_{C, \beta} \in \mathcal H} 4\beta c(C) \leq 2\epsilon\tau \max_{i} \sum_{p \in \mathcal P_i^h \cap \mathcal P_C} c(p)f_p \leq 2\epsilon \tau \sum_{e \in G_C} c(e) \max_{i} \sum_{p: e \in p, p \in \mathcal P_i^h} f_p 
        \leq 2\epsilon \tau \sum_{e \in G_C} c(e)x_e,
    \end{align*}
    where the last two inequalities follow respectively from the fact that all paths in $\mathcal P_C$ are contained in $G_C$, and from the constraints of \ref{lp:lcsf}. To conclude, note that $2\epsilon \tau = 2\log k$.
\end{proof}

\begin{lemma}
    \label{lem:lcsf_total_cost}
    The total weighted cost of candidate solutions is at most $\sum_{C, \beta} 4\beta c(H_{C, \beta}) \leq O\left(\frac{\log^2 k}\epsilon(\alpha + \log k)\right)X$.
\end{lemma}
\begin{proof} 
    By \Cref{lem:lcsf_charging,lem:lcsf_chunk_cost}, the weighted chunk cost is at most
    \[\sum_{H_{C, \beta} \in \mathcal H} 4\beta c(C) \leq 
    O(\log k) \sum_{e \in E} c(e) x_e |\{C: e \in G_C\}| \leq 
    O\left(\frac{\log^3 k}{\epsilon \log \tau}\right) \sum_{e \in E} c(e) x_e.\]
    Combining the chunk cost with the cost of rooted portions given by \Cref{lem:lcsf_rooted_cost}, we obtain 
    \[\sum_{C, \beta} 4\beta c(H_{C, \beta}) \leq 
    O\left(\alpha \frac {\log^2 k}\epsilon + \frac{\log^3 k}{\epsilon \log \tau} \right) \sum_{e \in E} c(e)x_e \leq 
    O\left(\frac {\log^2 k}{\epsilon} \left(\alpha + \frac{\log k}{\log \log k + \log (1/\epsilon)}\right)\right) \sum_{e \in E} c(e) x_e.\]
    Note that this analysis is with respect to the LP solution $(x, f)$ \emph{after} it has been scaled up by a factor $4/3$; this only changes the cost bound by a constant factor. 
\end{proof}
\noindent
Following the discussion above, this concludes the proof that there exists a low-density candidate, and therefore concludes the proof of \Cref{lem:lcsf_cost_main}. Along with \Cref{lem:lcsf_length}, we obtain \Cref{lem:lcst_to_lcsf_main}.

\section{Bicriteria Hardness of Approximations in General Graphs}\label{sec:hardlcsf}
\begin{figure}[H]
    \centering
    \begin{subfigure}[t]{0.3\textwidth}
        \centering
        \includegraphics[width=1\linewidth]{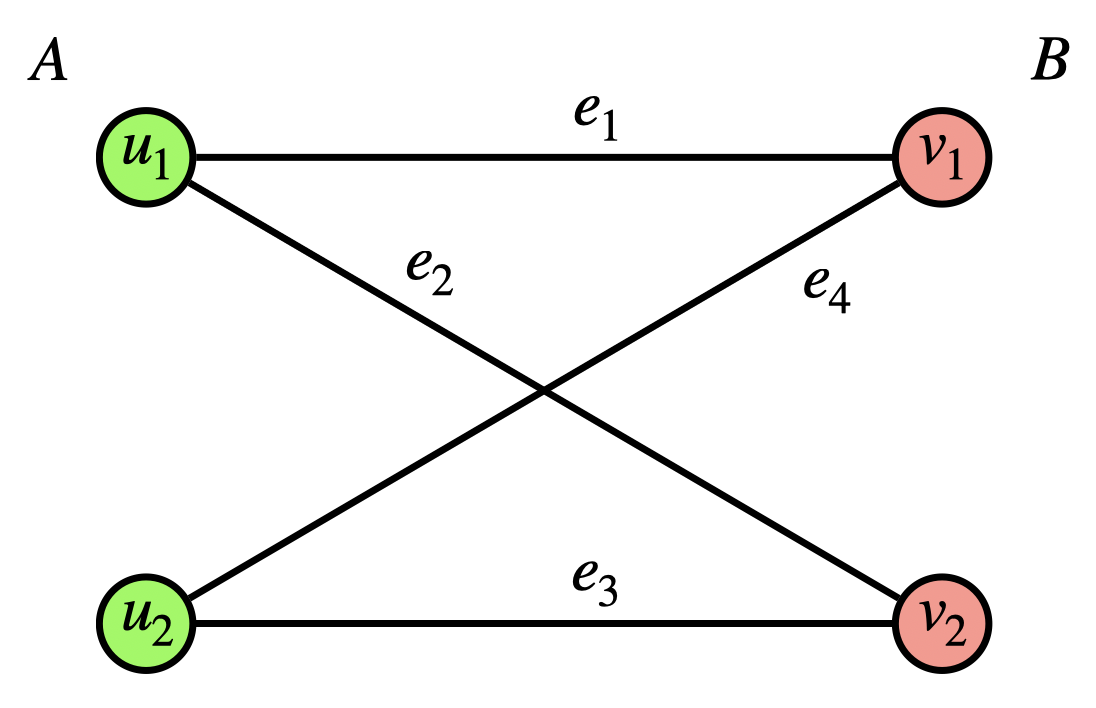}
        \caption{Min-Rep instance on $G$.}
        \label{subfig:hard1}
    \end{subfigure}%
    ~ 
    \begin{subfigure}[t]{0.7\textwidth}
        \centering
        \includegraphics[width=.99\linewidth]{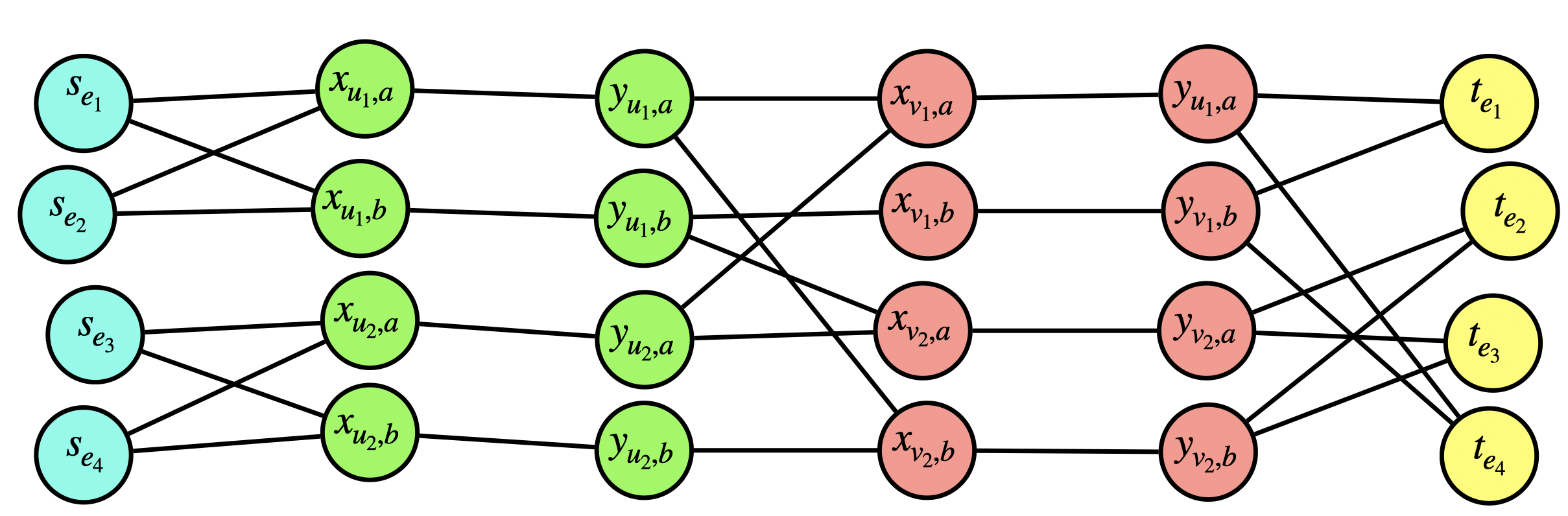}
        \caption{LC-SF instance on $H$.}
                \label{subfig:hard2}
    \end{subfigure}
    ~
    \begin{subfigure}[t]{\textwidth}
        \centering
        \includegraphics[width=.72\linewidth]{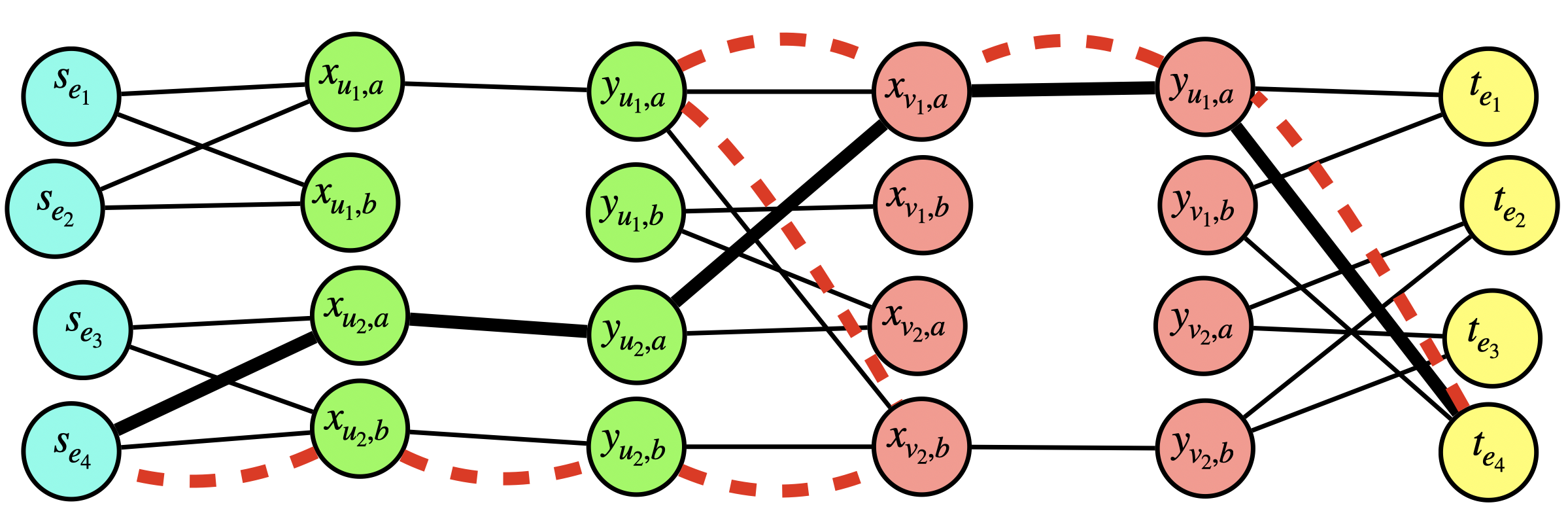}
        \caption{LC-SF solution $F$ in black edges from the Min-Rep instance on $G$ with labels $\Sigma_w=\{a,b\}$ for all $w\in A\cup B$ and valid label pairs $\Pi_{e_1}=\{(a,a),(b,b)\},\Pi_{e_2}=\{(a,b),(b,a)\},\Pi_{e_3}=\{(a,a),(b,b)\},\Pi_{e_4}=\{(a,a)\}$ and solution $S_{u_1}=\{a\},S_{u_2}=\{A,b\},S_{v_1}=\{a\},S_{v_2}=\{b\}$. In bolded black edges is a length $h$ path in $F$ from $s_{e_2}$ to $t_{e_2}$, and in dashed red edges is a cheaper but longer path from $s_{e_2}$ to $t_{e_2}$; removing the edge $(x_{u_2,a},y_{u_2,a})$ from $F$ would make it cheaper with length slack $3$.}
        \label{subfig:hard3}
    \end{subfigure}%
    \caption{The reduction from Min-Rep to LC-SF. \Cref{subfig:hard1}: a Min-Rep instance on graph $G=(A\cup B=\{u_1,u_2\}\cup\{v_1,v_2\},E=\{e_1,e_2,e_3\})$, with some unpictured set of labels and valid label pairs. \Cref{subfig:hard2}: the corresponding LC-SF instance $H$ where $V(H)_1$ is blue, $V(H)_2,V(H)_3$ are green, $V(H)_4,V(H)_5$ are red, and $V(H)_6$ is yellow. Note the reduction to LC-DSF is identical but with edges directed from left to right. \Cref{subfig:hard3}: an LC-SF solution in $H$ corresponding to a Min-Rep solution in $G$.}
    \label{fig:hard}
\end{figure}

In this section we reduce the Min-Rep problem to LC-SF and LC-DSF in general graphs. 
The input for Min-Rep includes a bipartite graph $G = (A \cup B, E)$, labels $\Sigma_{v}$ for each vertex $v \in A \cup B$, and for each edge $(u,v)\in E$ a non-empty relation $\Pi_{u,v} \subseteq \Sigma_{u} \times \Sigma_{v}$ consisting of the valid label pairs for $(u,v)$.
We must choose a subset of labels $S_{v} \subseteq \Sigma_{v}$ for all $v \in A \cup B$ such that every $(u,v) \in E$ is satisfied by a valid label pair, that is, there exist $a \in S_{u}$ and $b \in S_{v}$ such that $(a,b) \in \Pi_{u,v}$. The objective is to minimize $\sum_{v \in A \cup B} |S_{v}|,$ i.e.\ the total number of chosen labels across all vertices.
In particular, Min-Rep admits no $o\left(2^{\log^{1 - \epsilon}{n}}\right)$ approximation for all $\epsilon > 0$ under standard assumptions \cite{dinitz2015label}.
\thmHard*
\begin{proof}
    We describe a reduction from Min-Rep to LC-SF such that any solution on the LC-SF instance with length slack $\beta<3$ has cost no cheaper than the optimal length-exact solution. The reduction to LC-DSF will be very similar but exhibit a stronger hardness for length slack. 
    
    \paragraph{The Reduction.} Given a Min-Rep instance $G = (A \cup B, E), \left\{\Sigma_{v}\right\}, \left\{\Pi_{e}\right\}$ and $\beta<3$, we set a constant $\delta$ satisfying $0<\delta < \frac{3 - \beta}{4\beta}$ which will serve as the length for certain edges to get the length slack hardness guarantee, and construct an instance of LC-SF on graph $H$ as follows:
    \begin{itemize}
        \item \textbf{\textit{Vertices of $H$}}: for each $e \in E$ there is a source $s_{e}$ and sink $t_{e}$ which form our terminal pairs. For each $u \in A$ and $a \in \Sigma_{u}$ (resp.\ $v \in B$ and $b \in \Sigma_{v}$) we have vertices $x_{u,a}$ and $y_{u,a}$ (resp.\ $x_{v,b}$ and $y_{v,b}$).
        \item \textbf{\textit{Edges of $H$}}: $E_{H}$ contains the following edges.
            \begin{itemize}
                \item \textbf{\textit{Sources}}: for each $e=(u,v) \in E,a \in \Sigma_{u}$, there is an edge between $s_{e},x_{u,a}$ with cost $0$ and length $\delta$. 
                \item \textbf{\textit{Sinks}}: for each $e=(u,v) \in E,b \in \Sigma_{v}$, there is an edge between $y_{v,b},t_{e}$ with cost $0$ and length $\delta$.
                \item \textbf{\textit{Labels}}: for each $u \in A$ and $a \in \Sigma_{u}$ (resp.\ $v \in B$ and $b \in \Sigma_{v}$) there is an edge between $x_{u,a},y_{u,a}$ (resp.\ $x_{v,b},y_{v,b}$) with cost $1$ and length $\delta$. 
                \item \textbf{\textit{Longcuts}}: for each $(u,v) \in E$ and $(a,b) \in \Pi_{(u,v)}$, there is an edge between $y_{u,a},x_{v,b}$ with cost $0$ and length $1$.
        \end{itemize}
        \item \textbf{\textit{Terminal Pairs and Length Constraint}}: the terminal pairs is the set $\left\{\left(s_e,t_e\right):e\in E\right\}$ and each pair has length constraint $h := 1 + 4\delta$. 
    \end{itemize}
    Observe that $H$ is $6$-partite, as $V(H)$ can be partitioned into the following $6$ independent sets:
    \begin{align*}
        & V(H)_1:=\left\{s_e:e\in E\right\} & V(H)_2:=\left\{x_{u,a}:u\in A,a\in \Sigma_u\right\} 
        && V(H)_3:=\left\{y_{u,a}:u\in A,a\in \Sigma_u\right\}   \\ & V(H)_4:=\left\{x_{v,b}:v\in B,b\in \Sigma_v\right\} 
        & V(H)_5:=\left\{y_{v,b}:v\in B,b\in \Sigma_v\right\} &&  V(H)_6:=\left\{t_e:e\in E\right\}
    \end{align*}
    Furthermore, any edge in $H$ goes from $V(H)_i$ to $V(H)_{i+1}$ for some $i$: source edges go from $V(H)_1$ to $V(H)_2$, sink edges go from $V(H)_5$ to $V(H)_6$, label edges go from $V(H)_2$ to $V(H)_3$ or $V(H)_4$ to $V(H)_5$, and longcuts go from $V(H)_3$ to $V(H)_4$. The longcuts make it so that any path connecting a terminal pair has to give up at least one longcut, and if it takes multiple (free cost) longcuts to be cheaper it will give up huge length.
    \paragraph{Correctness of the Reduction.} It is clear that the reduction can be performed in polynomial time with respect to the input size. 
    
    First, we show that if there is a feasible Min-Rep solution with $C$ labels, then there exists a feasible LC-SF solution $F$ in the LC-SF instance defined earlier with cost $C$. Let $S_{A}$ and $S_{B}$ be the label assignments in the Min-Rep solution consisting of pairs $(u,a)$ (where $a \in \Sigma_{u}$) and $(v,b)$ (where $b \in \Sigma_{v}$), respectively. We have $|S_{A}| + |S_{B}| = C$. We choose $F$ as follows:
    $$ F:=\left\{ (x_{u,a},y_{u,a}): (u,a)\in S_A \right\}\cup \left\{ (x_{v,b},y_{v,b}):(v,b)\in S_B \right\} \cup \left\{ e \in E(H): c(e)=0\right\}$$
    The first and second sets in the union are the label edges corresponding to the valid label pairs chosen by the Min-Rep solution, and the third set is just taking every free-cost edge just because it can (which includes all the source/sink edges and longcuts). So the only edges of $H$ not chosen by $F$ are label edges, corresponding to the label pairs that the Min-Rep solution did not take. It is clear then by construction that $F$ has total cost $C$. 
    We next show that every terminal pair has a length $(1 + 4\delta)$ path in $F$. Fix a pair $(s_{e}, t_{e})$ corresponding to edge $e = (u,v)$. Since the Min-Rep solution is feasible, there exists $(a,b) \in \Pi_{u,v}$ such that $(u,a) \in S_{A}$ and $(v,b) \in S_{B}$. Thus, $F$ includes this edge, and $F$ contains the $s_e$-$t_e$ path $s_{e}$-$x_{u,a}$-$ y_{u,a}$-$x_{v,b}$-$y_{v,b}$-$t_{e}$, which has total length $\delta+\delta+1+\delta+\delta=1 + 4\delta = h$.
    
    We do the other direction and show that if there exists an LC-SF solution $F$ that connects each terminal pair with a path $P_{e}$ of length at most $\beta \cdot h$ (for $\beta < 3$), then $G$ has a valid Min-Rep solution of cost at most $c(F)$. For every $v\in A\cup B$ let
    $$ S_v:=\{f\in \Sigma_v:(x_{v,f},y_{v,f})\in F\} $$
    Since label edges are the only edges with nonzero cost, we have $\sum_v |S_v|=c(F)$. It suffices to show that $\{S_v\}_v$ is a feasible Min-Rep solution for $G$, by letting $F$ be an optimal LC-SF solution.
    
    We will show for every $e=(u,v)\in E$ its label satisfied by a label pair. Let $e\in E$ and $P_e\subseteq F$ be an $s_e$-$t_e$ path of length at most $l(P_{e}) \leq \beta (1 + 4\delta) < 3$ (which exists since $\delta < \frac{3 - \beta}{4\beta}$). Further let $V(H)_A:=V(H)_1\cup V(H)_2\cup V(H)_3,V(H)_B:=V(H)_4\cup V(H)_5\cup V(H)_6$; by construction the only edges crossing $V(H)_A,V(H)_B$ are the longcuts. Since $s_e\in V(H)_A,t_e\in V(H)_B$, the path $P_e$ must cross $V(H)_A,V(H)_B$ an odd number of times. If $P_e$ crosses $V(H)_A,V(H)_B$ more than once, it must take a longcut at least $3$ times; this is impossible since $l(P_e)<3$. Therefore, $P_e$ crosses $V(H)_A,V(H)_B$ exactly once, and in particular, if $F$ is an optimal solution then there is no cheaper LC-SF solution with length slack less than $3$.

    We observe some structure about $P_e$; in particular, we can break up $P_e$ into two subpaths $P_{e,A},P_{e,B}$ where $P_{e,A}$ is only in $V(H)_A$ and $P_{e,B}$ is only in $V(H)_B$, both only contain non-longcuts. Then $P_{e,A},P_{e,B}$ are connected by a longcut $(y_{u,a},x_{v,b})$, which exists only if $(a,b)\in\Pi_{u,v}=\Sigma_u\times\Sigma_v$. Moreover, the last edge in $P_{e,A}$ must be a label edge of the form $(x_{u,a},y_{u,a})$, implying $a\in S_u$ by definition of $S_u$. Symmetrically, the first edge in $P_{e,B}$ is a label edge $(x_{v,b},y_{v,b})$, implying $b\in S_v$. Therefore $e=(u,v)$ is satisfied by the (valid) label pair $(a,b)\in \Pi_{u,v}$ with $a\in S_u,b\in S_v$. This holds for any choice of $e\in E$, so the full collection $\{S_v\}_{v\in A\cup B}$ is a feasible Min-Rep solution for $G$.

    With the two directions above, we obtain the theorem statement. In particular, the second direction implies the same hardness for Min-Rep transfers to LC-SF even with length slack $\beta<3$.
        
    \paragraph{LC-DSF.} We do the same reduction but just direct edges in the direction from $V(H)_i$ to $V(H)_{i+1}$, and the same analysis follows analogously but for any positive length slack $\beta$ instead of for $\beta<3$, no path can take a longcut in the backwards direction to obtain a cheaper solution with longer length.
\end{proof}
\noindent See \Cref{fig:hard} for an illustration of the reduction.

\section{Acknowledgements}
\paragraph{AI Disclosure.} The authors initially discovered a poly-time LP-competitive $O\left(\frac{\log^4 k}{\epsilon^3}\right)$ approximation with $1+\epsilon$ length slack for LC-DST and a poly-time LP-competitive $O\left(\frac{\log^5 k\log n}{\epsilon}\right)$ approximation with $1+\epsilon$ length slack for LC-SF, both without AI assistance. We then gave these results to GPT-6 Astra. Through repeated interaction, GPT-6 Astra improved the LC-DST bound to $O\left(\frac{\log^2 k}{\epsilon} \right)$ and for LC-SF, GPT-6 Astra provided the path preprocessing and flow threshold arguments to remove the $O(\log n)$ in cost approximation, and the median trick to simplify the $1+ \epsilon$ length slack argument.
The authors then significantly cleaned up and conceptually simplified both results, including the framing of the LC-DST algorithm as charging cost-to-length and length-to-cost.  
The authors wrote the manuscript. GPT-6 Astra was used to proofread the manuscript, and to generate some figures from hand-drawn pictures. The authors verified the correctness of the results and take full responsibility for them.

\bibliographystyle{plainnat}
\bibliography{references}

\appendix

\section{Length-Constrained Directed Steiner Forest}\label{sec:lcdsf}
We use the modularity of the planar LC-DSF algorithm of \citet{chekuri2026lcdsf}.
\begin{lemma}[Theorem 4.1, Lemma 4.12, and Theorem 4.16 in \cite{chekuri2026lcdsf}]
    Suppose there is an $\alpha$ approximation with $\beta$ length slack for planar LC-DST w.r.t.\ \ref{lp:lcdst} that runs in time $\poly(n,1/\epsilon)$. Then there is an $\alpha\cdot\log^4k$ approximation with $6\beta$ length slack for planar LC-DSF that runs in time $\poly(n,1/\epsilon)$.
\end{lemma}
\noindent
Combining this with \Cref{thm:LCDST} improves the length slack for planar LC-DSF from $O(\log k)$ to $O(1)$.
\thmLCDSF*

\end{document}